\documentclass[twocolumn, aps,prx,groupedaddress,showpacs, superscriptaddress]{revtex4-1}
\usepackage{amsfonts}
\usepackage{amsmath}
\usepackage{amssymb}
\usepackage{txfonts}
\usepackage{pxfonts}
\usepackage{graphicx,bm,units,yfonts}
\usepackage{subcaption}
\usepackage[table]{xcolor}
\usepackage{hyperref}
\usepackage{movie15}

\newcommand{\CM}{{\mathbb C}}

\newcommand{\RM}{{\mathbb R}}
\newcommand{\SM}{{\mathbb S}}
\newcommand{\TM}{{\mathbb T}}
\newcommand{\ZM}{{\mathbb Z}}

\newcommand{\Oo}{{\mathcal O}}

\newcommand{\Ll}{{\mathcal L}}

\begin{document}

\title{Theory and Experimental Investigation of the Quantum Valley Hall Effect}

\author{Kai Qian}
\affiliation{Department of Physics, Yeshiva University, New York, NY, USA}

\author{David J. Apigo}
\author{Camelia Prodan}
\affiliation{Department of Physics, New Jersey Institute of Technology, Newark, NJ, USA}

\author{Yafis Barlas}
\author{Emil Prodan}
\affiliation{Department of Physics, Yeshiva University, New York, NY, USA}

\email{prodan@yu.edu}

\begin{abstract}
The quantum valley Hall effect (QVHE) has been observed in a variety of experimental setups, both quantum and classical. While extremely promising for applications, one should be reminded that QVHE is not an exact topological phenomenon and that, so far, it has been fully understood only qualitatively in certain extreme limits. Here we present a technique to relate QVHE systems with exact quantum spin-Hall insulators that accept real-space representations, without taking any extreme limit. Since the bulk-boundary correspondence is well understood for the latter, we are able to formulate precise quantitative statements about the QVHE regime and its robustness against disorder. We further investigate the effect using a novel experimental platform based on magnetically coupled spinners. Visual renderings, measurements and various tests of the domain-wall modes are supplied, hence giving an unprecedented insight into the effect.
\end{abstract}


\maketitle


\section{Introduction}

Graphene and related systems continue to be laboratories for new ideas and sources of remarkable new effects. With its low energy physics determined by two small pockets of the Brillouin zone, it led scientists to realize that a new effective observable, the valley, emerges in many physical situations. This observable can be controlled and manipulated like the spin \cite{GunawanPRL2006,GunawanPRB2006,RycerzNatPhys2007,XiaoPRL2007,YaoPRB2008} when the valley commutes with the dynamics of the low energy degrees of freedom. If the spectrum is gapped by breaking the inversion-symmetry of the honeycomb lattice,  a unique topological effect emerges \cite{XiaoPRL2007,FirozIslamCarbon2016} (see \cite{RenRPP2016} for a brief and informative update), the quantum valley-Hall effect (QVHE), akin to the quantum spin-Chern physics \cite{ShengPRL2006}. Here, we define the quantum spin-Chern insulators as time-reversal invariant systems with an additional $U(1)$ symmetry. The quantum valley-Hall insulators  ought to be defined the same way but the physical observable associated with the valleys has not been yet defined over the whole physical Hilbert space of the lattice, except for the continuum limit. This lack of a real-space representation made it difficult to quantify the bulk-boundary correspondence principle in these systems, though things are very clear in certain asymptotic limits. In contrast, for quantum spin-Chern insulators, the robustness against disorder and breaking of $U(1)$ symmetry is quantitatively well understood, in the sense that the conditions assuring the quantization of the spin-Chern numbers are explicitly and optimally known \cite{ProdanPRB2009}. Establishing a precise relation between the two effects is one of the goals of our work.

\vspace{0.2cm}

QVHE is quite appealing because it does not require breaking of the time-reversal symmetry or strong spin-orbit couplings. Instead, it can be engineered based on just a simple breaking of the inversion symmetry. It has been observed in many solid-state devices \cite{MakScience2014,GorbachevScience2014,SuiNatPhys2015,ShimazakiNatPhys2015,JuNature2015} and the interest continues to be strong, especially in the context of graphene bilayers \cite{MartinPRL2008,ZhangPRL2011,QiaoNL2011,WrightAPL2011,ZhangPNAS2013,PradaPRB2013,VaeziPRX2013,HuangArxiv2018}. It has been proposed in many photonic devices \cite{MaNJP2016,ChenArxiv2016,DongNatMat2017,ChenPRB2017,BleuPRB2017} and it was observed recently in a laboratory \cite{NohPRL2018}. QVHE has been enthusiastically embraced by the topological mechanics community, where there has been an explosion of laboratory demonstrations of the effect \cite{LuPRL2016,LuNatPhys2017,PalNJP2017,VilaPRB2017,ZhuArxiv2017,JiangArxiv2017,LiuPRA2018,WuSciRep2018,ChaunsaliPRB2018,ChernArxiv2018}. We were particularly motivated by the work of Ruzzene et al \cite{PalNJP2017,VilaPRB2017}, where the mechanical interface modes have been recorded in real time, providing a dramatic visual demonstration of signal guiding along zigzagged interfaces.

\vspace{0.2cm}

Even after so many successes, we need to keep in our minds that QVHE is not an exact topological effect. As opposed to the exact ones, QVHE systems are not surrounded by sharp phase boundaries and instead the effect can gradually fade way if the control over the parameters is lost. This is not a good news for practical applications because devices operate in real-world conditions and will be inherently subjected to stresses. If QVHE is to make its way into consumer products, one needs to identify the regimes where the effect occurs with certainty and to devise restoring processes if a material drifts away from those regimes.

\vspace{0.2cm}

In the present work, we use theory and experiment to add towards our body of knowledge about QVHE. One of the past theoretical difficulties was a lack of an exact topological insulator to be used as a reference system for the QVHE. Observing that a domain-wall configuration can be formally folded into a bilayered configuration with an edge, we are naturally led to a reference spin-Chern insulator accepting a real-space representation. Using the well understood bulk-boundary correspondence of the latter as well as the precise relation between the reference and the original QVHE system, we are able to identify general yet precise conditions in which QVHE occurs with certainty. We also identify a class of disorder perturbations which do not Anderson-localize the domain-wall states and, for generic disorder perturbations, we provide a quantitative estimator to asses the localization length of the domain-wall modes. These findings, the analysis behind it, as well as various ways to optimize the QVHE are presented in chapter~\ref{Sec:QuantQVHE}.

\vspace{0.2cm}

In chapter~\ref{Sec:ExpPlatform} we introduce a novel experimental platform based on magnetically coupled spinners \cite{ApigoArxiv2018}. Given that the relevant frequencies are within 20-40~Hz, the platform enables us to visualize and quantify the interface modes in a manner never achieved before. Aspects related to optimizing the QVHE regime and to the robustness of the interface modes against various perturbations can be tested first hand. Chapters~\ref{Sec:Bulk} and \ref{Sec:DomWall} present an in-depth analysis and measurements of the coupled spinners in both bulk and domain-wall configurations, respectively. In particular, section~\ref{Sec:ExpTechnique} discusses the details of our experimental measurements and supplies video recordings of the interface modes under various testing conditions. The main conclusions of our analysis are presented in chapter~\ref{Sec:Conclusions}.

\section{Quantitative theory of Valley Hall Effect}
\label{Sec:QuantQVHE}

In this chapter, we define an exact topological insulator, which is then used as a reference system to investigate the bulk-boundary principle in QVHE. Well established tools of analysis and the hallmarks of QVHE are reviewed first in sections~\ref{Sec:StrongBB} and \ref{Sec:Hallmarks}. The construction is given in section~\ref{Sec:RefSys} and the concrete analysis is contained in sections~\ref{Sec:Relation}, \ref{Sec:Existence} and \ref{Sec:Disorder}. The main statements are summarized in section~\ref{Sec:QVHERegime}.

\subsection{A look back at the Chern insulators}
\label{Sec:StrongBB}

The bulk-boundary correspondence principle for the 2-dimensional quantum Hall effect, hence also Chern insulators, was established by Hatsugai \cite{HatsugaiPRL1993}. The strong version of the principle, i.e. its extension to the context of irrational magnetic flux and disorder, has been proven by Kellendonk, Richter and Schulz-Baldes \cite{KRS-BRevMathPhys2002}. These works provide a topological argument for the existence of chiral edge states and for their robustness against Anderson localization, in a 2-dimensional topological system from class A of classification table \cite{SRFL2008,RSFL2010,Kit2009}. The extension of the argument to higher dimensions can be found in \cite{ProdanSpringer2016}.

\vspace{0.2cm}

 The topological argument is in fact extremely simple. Given a gapped bulk Hamiltonian $(H,G)$, with $G$ a connected component of $\RM \setminus {\rm Spec}(H)$, and its half-space version $\widehat H$, one constructs the unitary operator $\widehat U_B = e^{2 \pi i f(\widehat H)}$ using only the boundary states. For this, one chooses $f: \mathbb R \rightarrow \mathbb R$ to be any smooth function with variation only inside $G$ and taking the values 0/1 above/below this bulk gap. Indeed, by expanding in the eigen-system of $\widehat H$, one can convince himself that only the states with energy $\hat \epsilon$ inside $G$ contribute to $\widehat U_B -I$. These states are exponentially localized near the boundary, hence $\widehat U_B -I$ is one-dimensional in essence. Then the winding number:
\begin{equation}
{\rm Wind}(\widehat U_B) = \imath \, {\rm Tr}\big \{\widehat U_b [U_B,X_\|] \big \},
\end{equation}
written here in real space \cite{M-SSHPPRL2014}, is a topological invariant which obeys the equality \cite{HatsugaiPRL1993,KRS-BRevMathPhys2002}: \begin{equation}
{\rm Ch}(P_G) = {\rm Wind}(\widehat U_B), \quad P_G = \chi_{(-\infty,G]}(H).
\end{equation}
Throughout, $\chi_A$ will represent the characteristic function of the set $A$, $\bm X$ the position operator and $X_\|$ the position operator along the edge. Also, ${\rm Ch}$ refers to Chern number of a projection, which also has a real space representation \cite{BES1994}. Then, existence of disorder-immune edge modes follows from the fact that, if any parts of the boundary spectrum ${\rm Spec}(\widehat H) \cap G$ is Anderson localized, then $\widehat U_B$ can be constructed entirely from localized states, by properly adjusting the variation of $f$, in which case ${\rm Wind}(\widehat U_B)=0$. But this cannot happen if ${\rm Ch}(P_G) \neq 0$. This summarizes the strong bulk-boundary principle for Chern insulators. 

\vspace{0.2cm}

Our plan is to use these extremely effective tools and resolve the physics of the quantum valley Hall effect. They will not be used directly on such systems but rather on closely related reference exact topological systems.

\subsection{Hallmarks of the quantum valley Hall effect}
\label{Sec:Hallmarks}

Here we collect and comment on the characteristics of the quantum valley Hall effect, as they emerged over the years from both theoretical and experimental studies. Henceforth, we assume a gapped bulk Hamiltonian $(H,G)$ with two degrees of freedom per unit cell, assumed translational invariant for the beginning. We let $F$ stand for the Berry curvature of a gap projection $P_G=\chi_{(-\infty,G]}(H)$, which for translational invariant systems can be conveniently computed as \cite{AvronCMP1989}:
\begin{equation}\label{Eq:Curvature}
F(\bm k) = 2 \pi \imath \, {\rm tr} \bigg( P_G(\bm k) \big[ \partial_{k_{1}} P_G (\bm k),\partial_{k_{2}} P_G (\bm k) \big] \bigg).
\end{equation}
Above, ${\rm tr}$ is the trace over the two internal degrees of freedom. The main characteristics of QVHE are as follows:
\begin{itemize}
\item Time reversal is conserved. As such $F(\bm k) = - F(-\bm k)$ and necessarily $\int_{BZ} F(\bm k) {\rm d} \bm k =0$.
\item The are concentrations of the Berry curvature $F(\bm k)$ near two special $\bm K$ and $\bm K'=-\bm K$ quasi-momenta, which mark the valleys. They result from the splitting of a pair of Kramer degenerate Dirac nodes of the bulk  bands.

\item In the vicinities of the valleys: 
\begin{equation}
\int_{{\rm Vec}(\bm K)} F(\bm k) {\rm d}\bm k = -\int_{{\rm Vec}(\bm K')} F(\bm k) {\rm d}\bm k \simeq \frac{1}{2}.
\end{equation}

\item A pair of counter-propagating quasi-chiral energy bands emerge along an interface between the original system and its mirror reflection.

\item The quasi-chiral states are absent when the system is simply cut in half.

\item There is robustness of the interface modes against disorder but it is highly dependent on the orientation of the interface and the type of disorder.
\end{itemize}

A quantitative theory of QVHE will have to account for all these generic characteristics. Perhaps the most striking one is the emergence of the quasi-chiral modes along an interface but not along edges. Understanding the qualitative difference between these two cases is the starting point for our theory.

\subsection{An exact topological reference system}
\label{Sec:RefSys}

\begin{figure}
\includegraphics[width=\linewidth]{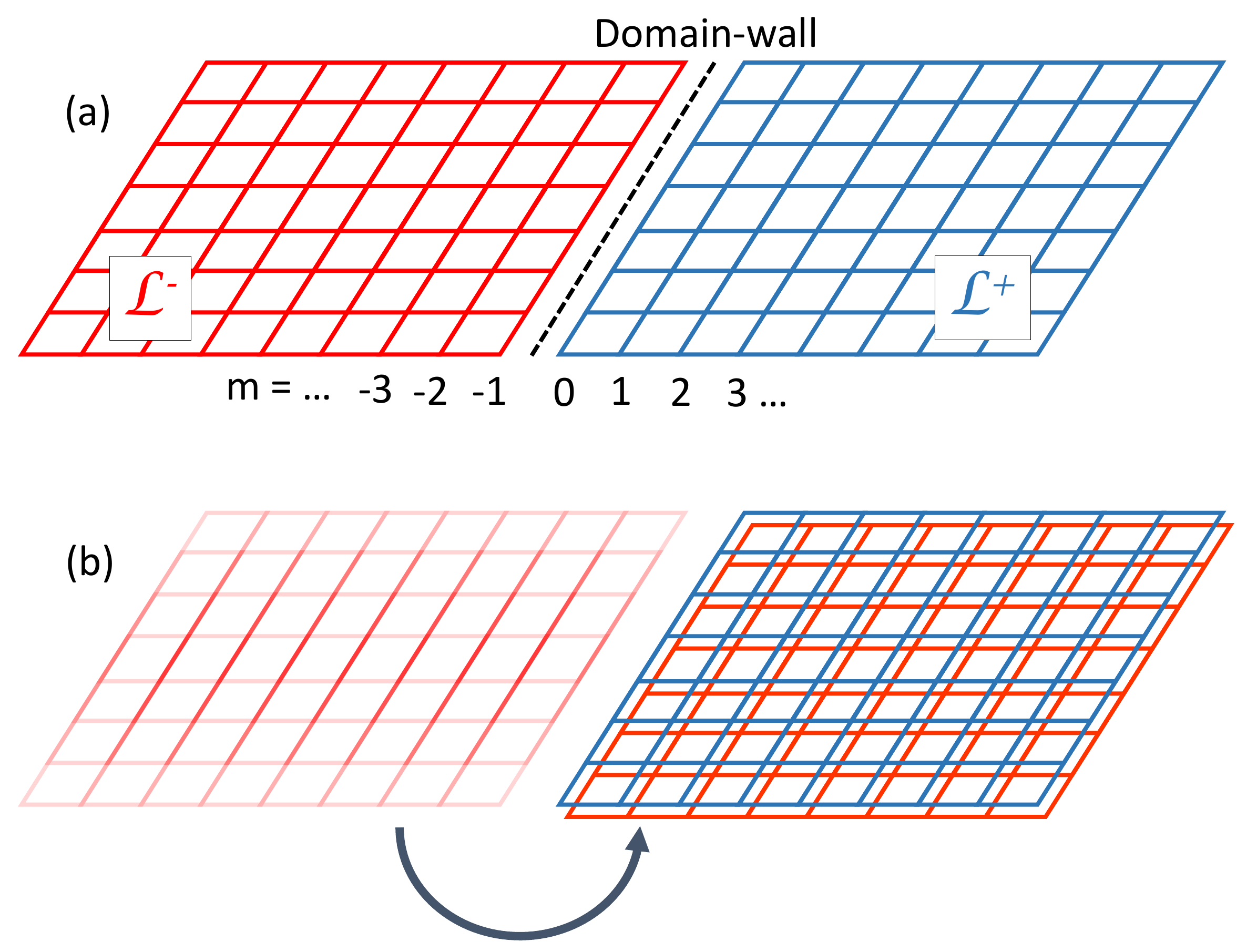}
\caption{\small {\bf Lattice folding.} (a) The domain-wall configuration. (b) Folding of the lattice into an edge configuration.}
\label{Fig:Th1}
\end{figure}

We assume here the presence of a domain wall. It is important to define the latter somewhat more precise. Henceforth, by a domain wall (DW) we understand a rapid spatial variation of one parameter of the Hamiltonian. This variation is entirely confined in a thin strip. Furthermore, we assume that the Hamiltonians on the two sides of the domain wall are related by a reflection symmetry. These are our broad working assumptions but, as we shall see, the QVHE regime is achieved in a more restrictive setting. Mapping those conditions is one of the main goals of our analysis.

\vspace{0.2cm}

We recall that, in the bulk-boundary correspondence, the bulk primitive cell needs to be adjusted to fit the interfaces or the edges (see {\it e.g} \cite{ProdanSpringer2016}[p.~48]). In the process, the primitive cell may become larger as it happens, for example, with the armchair edge of graphene. However, such enlargement is not beneficial to the QVHE because one wants to keep the valleys as separated as possible and expanding the unit cell leads to a folding of the Brillouin zone. Hence, we assume from the start that the DW is such that a choice of the bulk primitive cell can be made without enlarging the original primitive cell of the system. We also assume that a pair of shift operators $S_\|$ and $S_\bot$ have been chosen, of which $S_\|$ is along to the DW and $S_\bot$ crosses the wall, not necessarily perpendicularly (hence there are many options for $S_\bot$). With these choices, the system can be described on the Hilbert space $\mathbb C^2 \otimes \ell^2(\mathbb Z^2)$, with $S_\| |n,m\rangle = |n+1,m\rangle$ and $S_\bot |n,m\rangle = |n,m+1\rangle$. Hence, the coordinate $n$ runs along the DW, the center of the latter being assumed to be located between the primitive cells with labels $m=0$ and $m=-1$. All these will be exemplified in Section~\ref{Sec:DomWall} on a concrete example.

\begin{figure}
\includegraphics[width=0.9\linewidth]{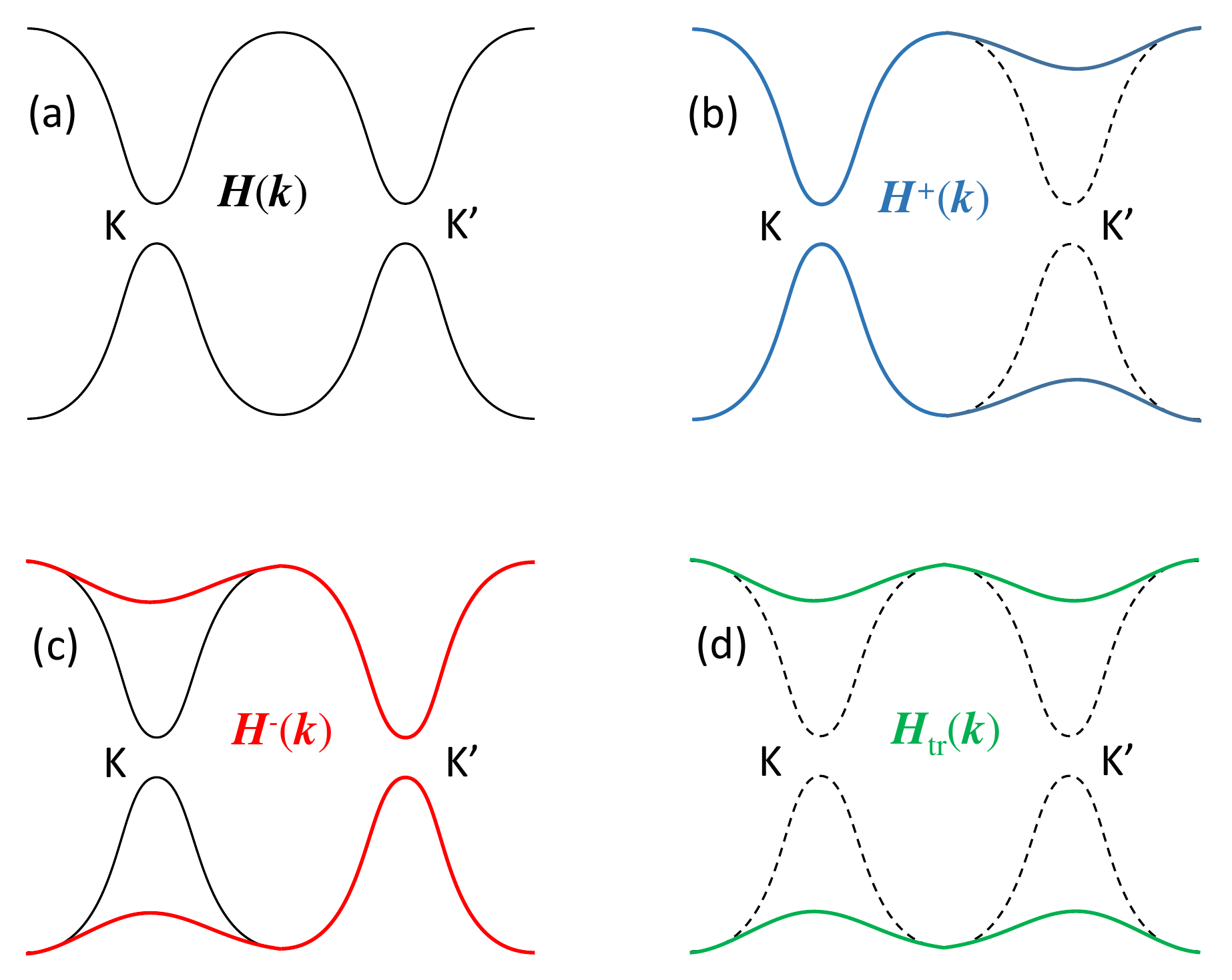}
\caption{\small {\bf Schematics of the continuous extensions of the Hamiltonians.} (a) The original Hamiltonian $\bm H$; (b,c) The extensions to $\bm H^\pm$, respectively; (d) The emerging trivial Hamiltonian $\bm H_{\rm tr}$. All Hamiltonians are represented by a section of their band spectral structure.}
\label{Fig:Th2}
\end{figure}

\vspace{0.2cm}

When a straight DW divides the lattice into two disjoint sectors $\ZM^2 = \Ll^- \cup \Ll^+$, as described above and illustrated in Fig.~\ref{Fig:Th1}(a), the most generic Hamiltonian for the DW configuration, excluding the disorder, takes the form:
\begin{equation}\label{Eq:OrigHam}
H=H_+ + H_- + V_{\rm DW},
\end{equation}
where:
\begin{equation}
H_+ = \sum_{\bm x,\bm x' \in \Ll^+} h_{\bm x -\bm x'} \otimes |\bm x\rangle \langle \bm x' |, \quad H_- = \mathcal I H_+ \mathcal I,
\end{equation}
and $V_{\rm DW}$ is a potential localized near the DW and invariant under translations along the wall. Above, $\mathcal I$ is the reflection operation relative to the DW center, which may have a non-trivial action on the internal degrees of freedom (see Sec.~\ref{Sec:TC}). Also, $h$'s are $2 \times 2$ matrices. A key innovation of our analysis is the formal folding of the lattice around the DW, as illustrated in Fig.~\ref{Fig:Th1}(b). It is formal because none of the hopping matrices is changed in the process and only a relabeling is performed. This relabeling defines a unitary transformation:
\begin{equation}
\Gamma: \CM^2 \otimes \ell^2(\ZM^2) \rightarrow \CM^4 \otimes \ell^2(\Ll^+),
\end{equation} 
which brings the Hamiltonian to the form:
\begin{equation}\label{Eq:HatOrigHam}
\widehat{\bm H} = \sum_{\bm x,\bm x' \in \Ll^+} \begin{pmatrix} h_{\bm x -\bm x'} & 0 \\ 0 & h_{\bm x -\bm x'} \end{pmatrix} \otimes |\bm x\rangle \langle \bm x' | + \bm V_{\rm E},
\end{equation}  
with $\bm V_{\rm E}= \Gamma W_{\rm DW} \Gamma^\dagger$, a potential that is localized near the edge. Two important achievements to notice:
\begin{itemize}
\item The DW configuration has been reduced to an edge configuration for the bulk Hamiltonian:
\begin{equation}
\bm H = \sum_{\bm x,\bm x' \in \ZM^2} \begin{pmatrix} h_{\bm x -\bm x'} & 0 \\ 0 & h_{\bm x -\bm x'} \end{pmatrix} \otimes |\bm x\rangle \langle \bm x' |,
\end{equation}
acting on $\CM^4 \otimes \ell^2(\ZM^2)$. The edge configuration is the setting in which the machineries of \cite{HatsugaiPRL1993,KRS-BRevMathPhys2002} have been developed, hence the folding trick enables us to extend these results to a DW configuration too. Otherwise, a direct analysis of the bulk-boundary principle for a DW configuration requires additional tools \cite{KotaniJMP2014}. 
\item Because the folded bulk system consists of two identical copies of the original system, in the vicinity of the valleys:
\begin{equation}
\int_{{\rm Vec}(\bm K)} \bm F(\bm k) {\rm d}\bm k = -\int_{{\rm Vec}(\bm K')} \bm F(\bm k) {\rm d}\bm k \simeq 1,
\end{equation}
where, this time, $\bm F$ is the Berry curvature associated with the gap projection $\bm P_G = \chi_{(-\infty,G]}(\bm H)$.
\end{itemize}

\begin{figure}
\includegraphics[width=0.75\linewidth]{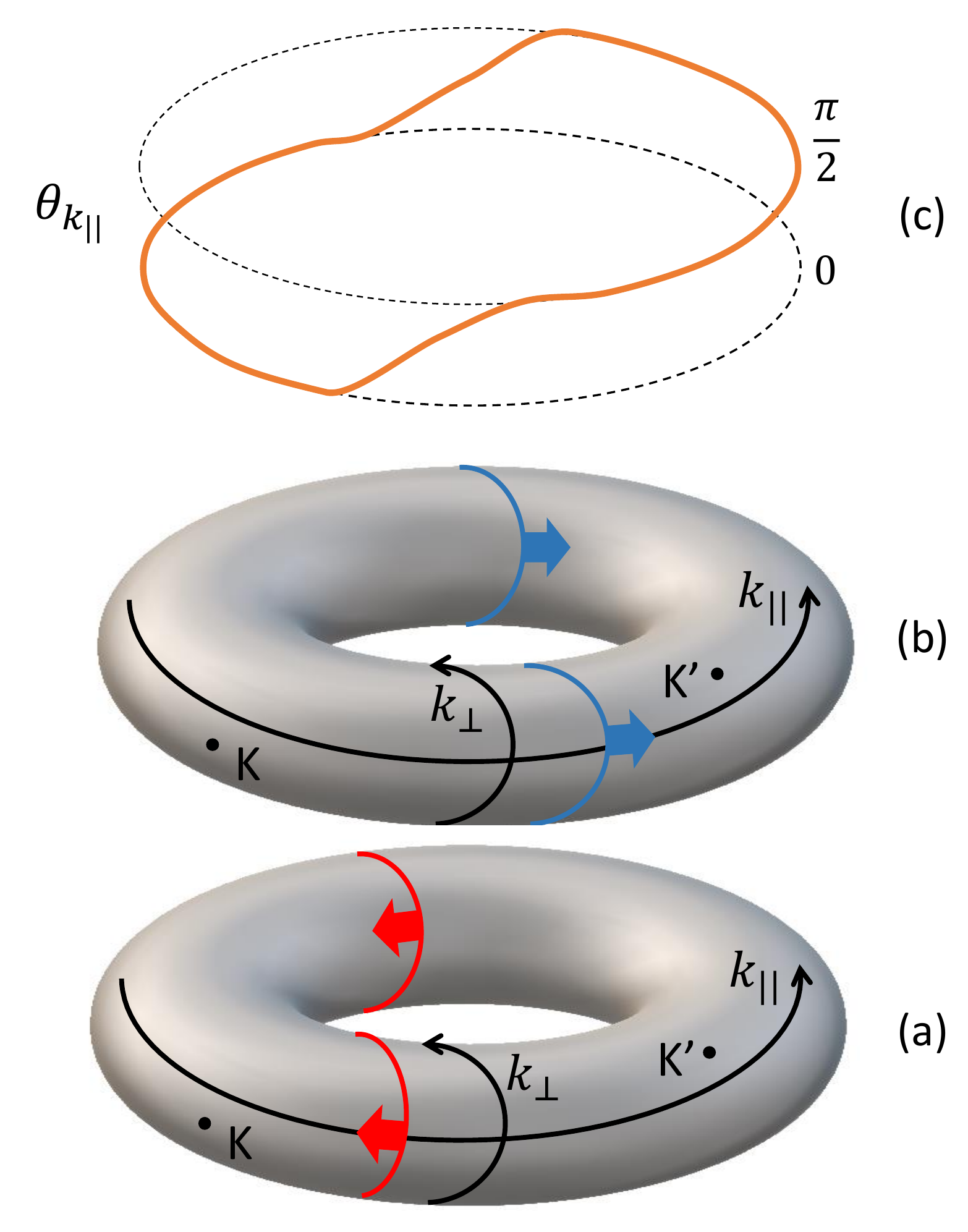}
\caption{\small {\bf The smooth extensions across the Brillouin torus.} (a) $\bm H(\bm k)$ is smoothly deformed from and in the direction of the blue arrows into what becomes $\bm H^+(\bm k)$. (b) $\bm H(\bm k)$ is smoothly deformed from and in the direction of the red arrows into what becomes $\bm H^+(\bm k)$. The color coding is the same as in Fig.~\ref{Fig:Th2}. It is important to note that the arrows are aligned with $k_\|$ and that there is an overlap between $\pm$ extensions, which in the text is called $\Oo$. (c) The angle $\theta_{k_\|}$ entering in Eq.~\ref{Eq:U}.}
\label{Fig:Th3}
\end{figure}

\vspace{0.2cm}

The last observation reflects one of the main differences between the simple edge and DW configurations and another innovation of our analysis is to take full advantage of  it. Indeed, since now the Berry curvature near a valley integrates to an integer, there is no topological obstruction against smoothly extending the gap projection $\bm P_G(\bm k)$  from ${\rm Vec}(K)$ to the entire Brillouin torus without adding any more Berry curvature flux. We will call this continuation $\bm P^+_G(\bm k)$. Note that $1- \bm P^+_G(\bm k)$ provides a similar extension of the spectral projection onto the upper spectrum. The conclusion at this point is that $\bm H(\bm k)$ can be continuously extended over the whole Brillouin torus to a topological Chern Hamiltonian $\bm H^+(\bm k)$ such that the two coincide over more than half of the Brillouin torus. This is schematically shown in Fig.~\ref{Fig:Th2}(b). Similarly, an extension $\bm H^-(\bm k)$ is obtained if the argument is repeated for the valley $K'$, which is schematically shown in Fig.~\ref{Fig:Th2}(c). Let us state explicitly that for these Hamiltonians:
\begin{equation}\label{Eq:Ch1}
{\rm Ch}(\bm P^\pm_G) = \pm 1.
\end{equation}
We have been careful to choose these extensions so that there is an overlap region $\Oo$ where $\bm H^\pm(\bm k)$ coincide, such that we can actually construct a third Hamiltonian $\bm H_{\rm tr}(\bm k)$ as illustrated in Fig.~\ref{Fig:Th2}(d). Fig.~\ref{Fig:Th3} shows how these extensions are precisely performed over the Brillouin torus.

\vspace{0.2cm}

The above procedure leads us naturally to the reference topological Hamiltonian:
\begin{equation}\label{Eq:RefHam}
\bm H_{SC}(\bm k) = \begin{pmatrix} \bm H^+(\bm k) & 0 \\ 0 & \bm H^-(k) \end{pmatrix},
\end{equation}
which is a spin-Chern insulator, {\it i.e.} a time-reversal symmetric system with an additional $U(1)$ symmetry, the latter being generated by:
\begin{equation}
\Sigma(\bm k) = \begin{pmatrix} I_{4 \times 4} & 0 \\ 0 & -I_{4 \times 4} \end{pmatrix}.
\end{equation} 
An important observation is that the reference Hamiltonian is defined over the entire Brillouin torus and, as such, it admits a real-space representation where real physical edges can be considered. As long as the $U(1)$ symmetry is present, the two non-trivial Chern sectors of $\bm H_{\rm SC}$ decouple and the strong version of the bulk-boundary described in \ref{Sec:StrongBB} applies separately on these sectors. As such, a pair of counter-propagating boundary chiral modes emerges when the Hamiltonian \eqref{Eq:RefHam} is halved in such a way that $U(1)$ symmetry is preserved.

\subsection{Relation with original system}
\label{Sec:Relation}

In \eqref{Eq:RefHam}, the $(\pm)$ Hamiltonians act on separate 4-dimensional internal spaces. We can make them act on the same internal space by using  a unitary transformation that rotates the 8-dimensional internal space of $\bm H_{\rm SC}$:
\begin{equation}\label{Eq:U}
\bm U(\bm k) = \begin{pmatrix}  \cos \theta_{k_\|} \, I_{4 \times 4} & - \sin \theta_{k_\|}  \, I_{4 \times 4} \\ \sin \theta_{k_\|}  \, I_{4 \times 4} & \cos \theta_{k_\|}  \, I_{4 \times 4} \end{pmatrix},
\end{equation}
with the angle $\theta$ specified in Fig.~\ref{Fig:Th3}. Then it is straightforward to see that:
\begin{equation}\label{Eq:MainId}
\bm H_{\rm QVHE}(\bm k) = \bm U(\bm k) \, \bm H_{\rm SC}(\bm k) \, \bm U(\bm k)^\dagger
\end{equation} 
coincides with the original system augmented by the trivial Hamiltonian, that is:
\begin{equation}\label{Eq:QVHE}
\bm H_{\rm QVHE} (\bm k) = \begin{pmatrix} \bm H(\bm k) & 0 \\ 0 & \bm H_{\rm tr}(k) \end{pmatrix}.
\end{equation}
Note that the remarkable identity in \eqref{Eq:MainId} holds in part because $\bm H^\pm(\bm k)$ coincide with $\bm H(\bm k)$ on the zone $\Oo$ where $\theta$ has its variations. Obviously, $\bm H_{\rm SH} (\bm k)$and $\bm H_{\rm QVHE} (\bm k)$ have identical band spectra.

\vspace{0.2cm}

Although \eqref{Eq:QVHE} implies just a stacking of a trivial system on top of the system we want to study, this action is not benign because in many instances stacking and de-stacking can have non-trivial effects. This aspect will be analyzed in detail next.

\subsection{Existence of quasi-chiral edge bands}
\label{Sec:Existence}

Recall that any continuous family of matrices $N\times N$ matrices $M(\bm k)$ defined over the {\it whole} Brillouin torus can be Fourier decomposed:
\begin{equation}
M(\bm k) = \sum_{\bm q} m_{\bm q} e^{i \bm q \cdot \bm k}
\end{equation}
and then written as an operator on the real space:
\begin{equation}
M= \sum_{\bm q} m_{\bm q} \otimes S_{\|}^{q_1}S_{\bot}^{q_2}.
\end{equation}
We will use this standard procedure to transfer all Hamiltonians defined so far, as well as $\bm U(\bm k)$, from the quasi-momentum to the real-space representation. The real-space representation will bear the same symbols but the $\bm k$ dependence will, obviously, be dropped. 

\vspace{0.2cm}

Let $\Pi_+: \ell^2(\ZM^2) \rightarrow \ell^2(\Ll^+)$ be the isometry:
\begin{equation}
\Pi_+ |\bm x \rangle = \chi_{\Ll^+}(\bm x) |\bm x \rangle, \quad \bm x \in \ZM^2.
\end{equation}
Then $\widehat M = \Pi_+ M \Pi_+^\dagger$ is the half-space version of the operators, with abrupt Dirichlet boundary condition. Throughout we will use the hat to indicate this process. Now, due to the very particular way we performed the extensions as well as we constructed $\bm U$, we have:
\begin{equation}
\widehat{\bm H}_{\rm QVHE} = \Pi_+ \bm U  \, \bm H_{\rm SC} \, \bm U^\dagger \Pi_+^\dagger = \widehat{\bm U} \, \widehat{\bm H}_{\rm SC} \, \widehat{\bm U}^\dagger.
\end{equation}
Since $\widehat{\bm U}$ remains an unitary operator and conjugation by unitaries preserves the spectra, it follows that $\widehat{\bm H}_{\rm QVHE}$ and $\widehat{\bm H}_{\rm QVHE}$ are iso-spectral. Furthermore, in the absence of disorder, we can perform a Bloch-Floquet transformation with respect to $k_\|$ and automatically obtain:
\begin{equation}
\widehat{\bm H}_{\rm QVHE}(k_\|) = \widehat{\bm U}(k_\|) \, \widehat{\bm H}_{\rm SH}(k_\|) \, \widehat{\bm U}^\dagger(k_\|).
\end{equation}
We reached one of the main conclusions of our analysis, namely, that $\widehat{\bm H}_{\rm QVHE}$ and $\widehat{\bm H}_{\rm SH}$ have identical edge spectra. Since the latter displays a pair of chiral edge bands, we can conclude that $\widehat{\bm H}_{\rm QVHE}$ does too. Furthermore, the pair of chiral edge bands of $\widehat{\bm H}_{\rm SC}(k_\|)$ cannot be destroyed by any change of boundary condition which preserves the $U(1)$ symmetry, {\it e.g.} of the type:
\begin{equation}\label{Eq:ChB}
\widehat{\bm H}_{\rm SC}(k_\|) \rightarrow  \widehat{\bm H}_{\rm SC}(k_\|) + \begin{pmatrix} \bm V_{\rm E}(k_\|) & 0 \\ 0 & \bm V_{\rm E}(k_\|) \end{pmatrix},
\end{equation}
with $\bm V_{\rm E}(k_\|)$ localized near the edge. Since the diagonal Hamiltonians commute with $\widehat{\bm U}(k_\|)$:
\begin{align}
 \widehat{\bm U}(k_\|) \, & \left [ \widehat{\bm H}_{\rm SH}(k_\|) + \begin{pmatrix} \bm V_{\rm E}(k_\|) & 0 \\ 0 & \bm V_{\rm E}(k_\|) \end{pmatrix} \right ]\, \widehat{\bm U}^\dagger(k_\|) \\ \nonumber
=& \widehat{\bm H}_{\rm QVHE}(k_\|) + \begin{pmatrix} \bm V_{\rm E}(k_\|) & 0 \\ 0 & \bm V_{\rm E}(k_\|) \end{pmatrix},
\end{align}
and we can conclude that the pair of chiral edge bands of $\widehat{\bm H}_{\rm QVHE}$ also cannot be destroyed by such changes of boundaries conditions.

\begin{figure}
\includegraphics[width=0.75\linewidth]{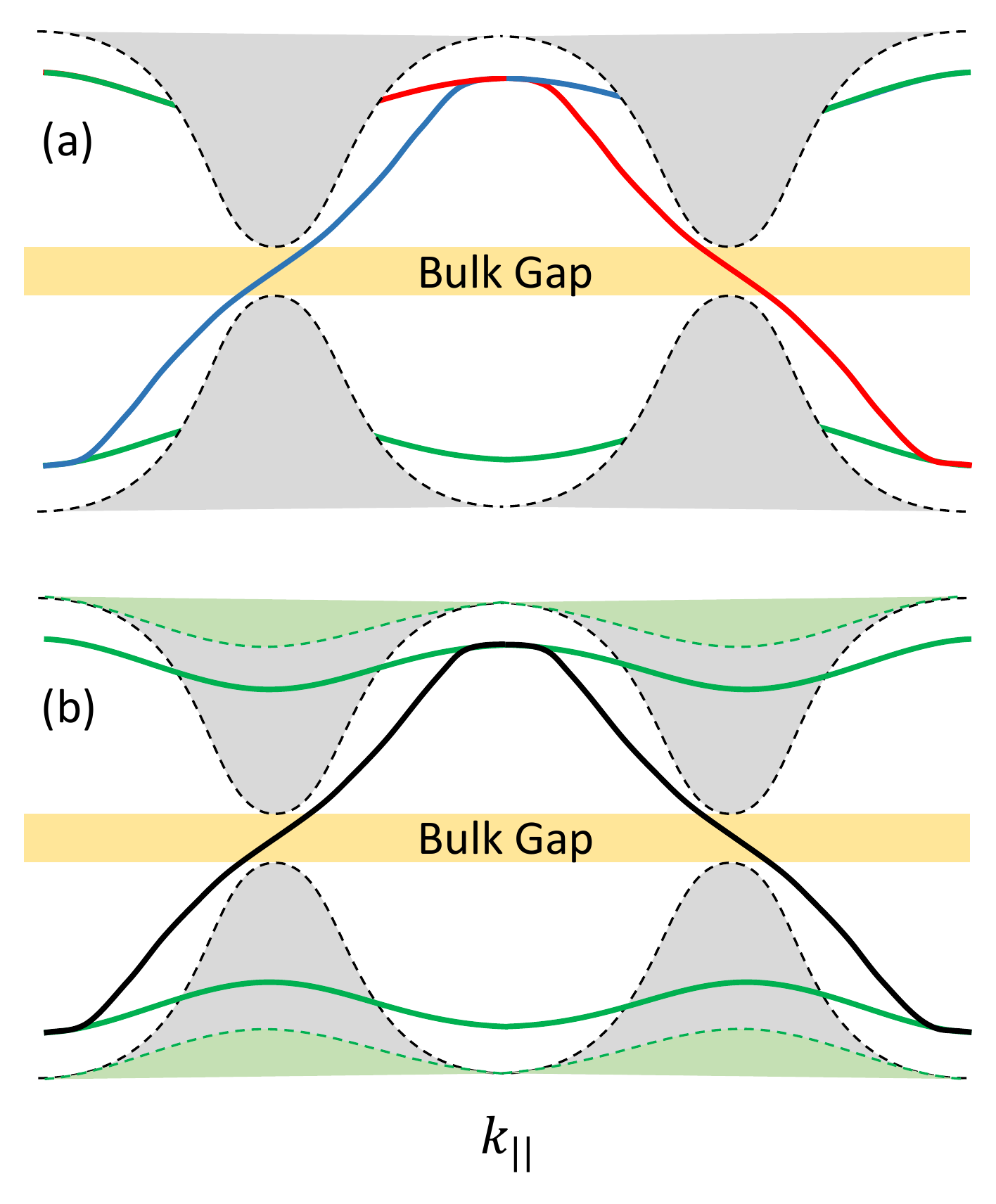}
\caption{\small {\bf Schematics of the edge spectra.} (a) The topological edge spectrum of the spin-Chern Hamiltonian $\bm H_{\rm SC}$ consists of a pair of counter-propagating chiral bands, shown in blue and red. (b) The topological edge spectrum coincides with the reunion of the spectra of the original Hamiltonian $\bm H$, shown in black, and of the trivial Hamiltonian $\bm H_{\rm tr}$, shown in green. The shaded areas represent the bulk spectra.}
\label{Fig:Th4}
\end{figure}

\vspace{0.2cm}

We investigate now the spectra in more details. Recall that our ultimate goal is to produce a statement about the edge spectrum of the physical Hamiltonian $\widehat{\bm H}$ and, so far, we only have a statement about $\widehat{\bm H}_{\rm QVHE}$. For this, we will use the following facts already established above:
\begin{itemize}
\item $\widehat{\bm H}_{\rm QVHE}$ displays a pair of mirror-imaged chiral edge bands (under $k_\| \rightarrow -k_\|$).
\item This pair of edge bands also coincide with the reunion of the edge spectra of $\widehat{\bm H}$ and $\widehat{\bm H_{\rm tr}}$. Recall that these two Hamiltonians coincide over the section $\Oo$ of the Brillouin torus.
\item The Hamiltonian $\bm H_{\rm tr}$ is topologically trivial, hence there are boundary conditions under which its edge spectrum consists of impurity bands residing very close to its bulk spectrum.
\end{itemize}

\vspace{0.2cm}

Based on these guiding facts, for those special boundary conditions, we arrive at the spectra shown in Fig.~\ref{Fig:Th4}. They give a simple explanation of the existence of the pair of edge bands of $\widehat{\bm H}$ crossing its bulk gap near $K$ and $K'$. Indeed, since the edge bands of $\widehat{\bm H}_{\rm tr}$ are very close to its bulk spectrum, the only way $\widehat{\bm H}_{\rm QVHE}$ can display a pair of chiral bands is if the edge bands of $\widehat{\bm H}$ cross its bulk gap as shown in Fig.~\ref{Fig:Th4}(b).

\vspace{0.2cm}

The remaining piece of the puzzle is understanding when is the DW configuration, as introduced at the beginning of section~\ref{Sec:RefSys}, leading to the special boundary conditions mentioned above. The spectrum of $\widehat{\bm H}_{\rm tr}$ is constrained on $\Oo$ to equal that of $\widehat{\bm H}$ and in rest it can be engineered to virtually take any values. As such, we only need to address the issue for $k_\| \in \Oo$. Then it is quite evident that those special boundary conditions occur whenever the original physical Hamiltonian $H$ from \eqref{Eq:OrigHam} is close enough to a reference translational symmetric Hamiltonian $H_0$. Indeed, then $\widehat{\bm H}$ in \eqref{Eq:HatOrigHam} is a deformation of $\widehat{\bm H}_0 = \Gamma H_0 \Gamma^\dagger$ and the latter displays only bulk spectrum. As such, the edge spectrum of $\widehat{\bm H}$ in the region $\Oo$, hence also of $\widehat{\bm H}_{\rm tr}$, must be as depicted in Fig.~\ref{Fig:Th4}. Let us point out that our arguments work for as long as this spectrum does not touch the bulk gap of $H$, hence the deformations from reference Hamiltonian $H_0$ need not be small. 

\subsection{The effect of disorder}
\label{Sec:Disorder}

We now introduce disorder:
\begin{equation}\widehat{\bm H} \rightarrow \widehat{\bm H} +\widehat{\bm V}_\omega,
\end{equation}
and, first, we identify a class of disorder perturbations which do not Anderson localize the edge spectrum. For this, we make a connection with the spin-Chern system, where the tools mentioned in \ref{Sec:StrongBB} assure us that the edge spectrum remains delocalized as long as the disorder preserves $U(1)$ symmetry (see \cite{ProdanJMP2009} for an explicit analysis). To establish the connection, we augment again with the trivial Hamiltonian:
\begin{equation}
\begin{pmatrix} \widehat{\bm H} + \widehat{\bm V}_\omega & 0 \\ 0 & \widehat{\bm H}_{\rm tr} + \widehat{\bm V}_\omega \end{pmatrix} = \widehat{\bm H}_{\rm QVHE} + \begin{pmatrix} \widehat{\bm V}_\omega & 0 \\ 0 & \widehat{\bm V}_\omega \end{pmatrix},
\end{equation}
and observe that:
\begin{align}
\widehat{\bm U}^\dagger \left [ \widehat{\bm H}_{\rm QVHE} + \begin{pmatrix} \widehat{\bm V}_\omega & 0 \\ 0 & \widehat{\bm V}_\omega \end{pmatrix} \right ] \widehat{\bm U}   = \widehat{\bm H}_{\rm SC} + \widehat{\bm W}_\omega, 
\end{align}
with $\widehat{\bm W}_\omega$ equal to:
\begin{equation}\label{Eq:WQ}
\tfrac{1}{2} \begin{pmatrix} \widehat{\bm R} \widehat{\bm V}_\omega \widehat{\bm R}^\dagger + \widehat{\bm R}^\dagger \widehat{\bm V}_\omega \widehat{\bm R} & -\imath ( \widehat{\bm R} \widehat{\bm V}_\omega \widehat{\bm R}^\dagger - \widehat{\bm R}^\dagger \widehat{\bm V}_\omega \widehat{\bm R}) \\ \imath (\widehat{\bm R} \widehat{\bm V}_\omega \widehat{\bm R}^\dagger - \widehat{\bm R}^\dagger \widehat{\bm V}_\omega \widehat{\bm R}) & \widehat{\bm R} \widehat{\bm V}_\omega \widehat{\bm R}^\dagger + \widehat{\bm R}^\dagger \widehat{\bm V}_\omega \widehat{\bm R} \end{pmatrix}. 
\end{equation}
Above, $\widehat{\bm R}$ stands for the translational invariant unitary operator whose Bloch decomposition coincides with $e^{\imath \theta_{k_\|}} \times I_{4 \times 4}$. It is important to notice that $\widehat{\bm R}$ is not conditioned by the fictitious extensions since the choice of $\theta_{k_\|}$ is made entirely based on the values of the Berry curvature $\bm F(\bm k)$. From \eqref{Eq:WQ}, we can see explicitly that the $U(1)$ symmetry is preserved if:
\begin{equation}\label{Eq:DisorderCond}
\widehat{\bm R} \widehat{\bm V}_\omega \widehat{\bm R}^\dagger = \widehat{\bm R}^\dagger \widehat{\bm V}_\omega \widehat{\bm R} \ \Leftrightarrow \ [\widehat{\bm V}_\omega,\widehat{\bm R}^2]=0.
\end{equation}
Every disorder perturbation satisfying \eqref{Eq:DisorderCond} cannot Anderson-localize the edge spectrum of $\widehat{\bm H}_{\rm CS}$, hence also of $\widehat{\bm H}_{\rm QVHE}$. The latter automatically implies that the edge spectrum of $\widehat{\bm H}$ remains delocalized under such perturbations.

\begin{figure}
\includegraphics[width=0.8\linewidth]{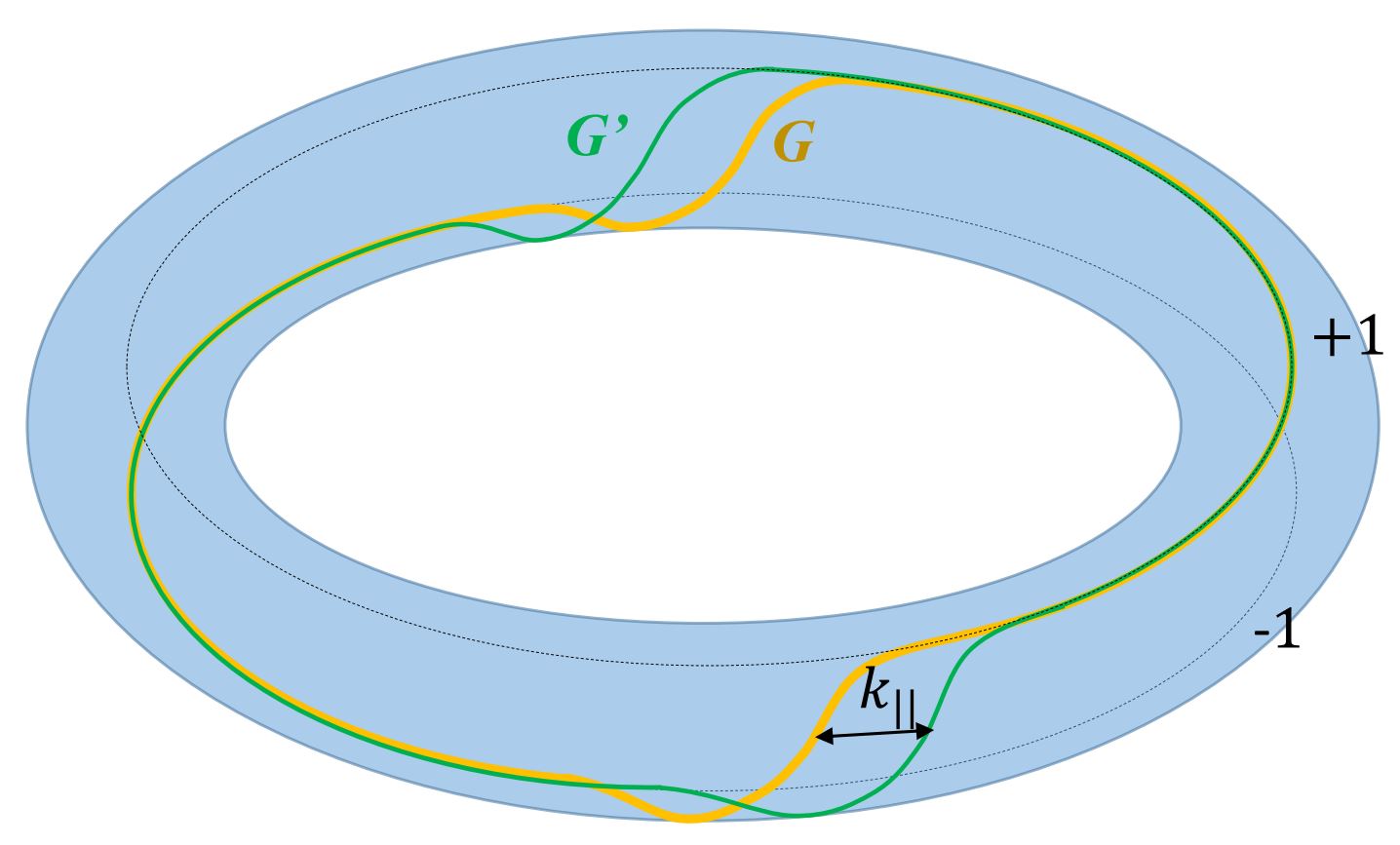}
\caption{\small {\bf The function $G: \SM^1 \rightarrow \SM^1$ entering the analysis of disorder.} It takes the constant values $\pm 1$ on two regions and in between it has a smooth variation. The function $G'$, shown in green, is related by $G'(z)=G(e^{\imath k_\|}z)$. The two functions appear together for the first time in our analysis in Eq.~\ref{Eq:Comm}.}
\label{Fig:Th5}
\end{figure}

\vspace{0.2cm}

The first important question is if \eqref{Eq:DisorderCond} has any solutions other than translational invariant $V_\omega$'s, and the answer is yes. Indeed:
\begin{equation}
\widehat{\bm R}^2 = G(S_\|),
\end{equation}
with $G: \SM^1 \rightarrow \SM^1$ having a profile as shown in Fig.~\ref{Fig:Th5}. Here and below, the unit circle $\SM^1$ represents the locus $|z|=1$ of the complex plane. We recall that if an operator $A$ commutes with an operator $B$, then it commutes with the entire algebra generated by the latter. But, due to the particular profile of function $G$, the algebra generated by $G(S_\|)$ is strictly smaller than the algebra generated by $S_\|$, hence there are indeed solutions to \eqref{Eq:DisorderCond} for which $[\widehat{\bm V}_\omega,S_\|] \neq 0$. These solutions very likely involve long-range bond disorder.

\vspace{0.2cm}

The second important question is what the above analysis tells us about on-site disorder potentials. While this type of potential do not satisfy the condition \eqref{Eq:DisorderCond}, we have a quantitative way to asses when this condition is at least approximately satisfied. Indeed, let:
\begin{equation}
V(\bm X) = \int_{\TM^2}{\rm d}\bm k \,  v(\bm k) e^{\imath \bm k \cdot \bm X}
\end{equation}
be an on-site potential. Recalling that $S_\|^\dagger X_\| S_\| = X_\|+1$, we have $e^{\imath \bm k \cdot \bm X} S_\| = e^{\imath k_\|}S_\| e^{\imath \bm k \cdot \bm X}$, hence:
\begin{equation}
V(\bm X) G(S_\|) = \int_{\TM^2}{\rm d}\bm k \,  v(\bm k) G(e^{\imath k_\|}S_\|)e^{\imath \bm k \cdot \bm X},  
\end{equation}
and:
\begin{equation}\label{Eq:Comm}
[V(\bm X),\widehat{\bm R}^2] = \int_{\TM^2}{\rm d}\bm k \,  v(\bm k) \big [G(e^{\imath k_\|}S_\|)-G(S_\|) \big ]e^{\imath \bm k \cdot \bm X} . 
\end{equation}
Inside the square brackets, one sees the function: 
\begin{equation}\label{Eq:F}
F_{k_\|}(z)=G(e^{\imath k_\|}z) - G(z)
\end{equation}
being applied to $S_\|$. The norm of the commutator \eqref{Eq:Comm} can be bounded by:
\begin{equation}
\big \|[V(\bm X),\widehat{\bm R}^2] \big \| \leq \int_{\TM^2}{\rm d}\bm k \,  |v(\bm k)| \, \sup_{z\in \SM^1} \big | F_{k_\|}(z) \big |
\end{equation}
Note that the first term in \eqref{Eq:F} is a function $G':\SM^1 \rightarrow \SM^1$ with the profile give by that of $G$ rotated by $k_\|$, as illustrated in Fig.~\ref{Fig:Th5}. It then becomes clear that the commutator \eqref{Eq:Comm} is small whenever the support of $v(\bm k)$ is concentrated near $k_\|=0$ and the variation of $G$ is slow.

\vspace{0.2cm}

The latter finding is in line with the now well accepted fact that the DW modes in QVHE are protected against random perturbations whose variations occur at length-scales larger than that of the lattice (see {\it e.g.} \cite{ZhangPNAS2013}). What our analysis adds new is a quantitative way to asses the effects of such potentials. For example, we propose as a useful estimator the ratio:
\begin{equation}\label{Eq:R}
R=\frac{\big \| \widehat{\bm V}_\omega - (\widehat{\bm R}^\dagger)^2 \widehat{\bm V}_\omega \widehat{\bm R}^2 \big \|}{\big \| \widehat{\bm V}_\omega + (\widehat{\bm R}^\dagger)^2 \widehat{\bm V}_\omega \widehat{\bm R}^2 \big \|},
\end{equation}
 between the off-diagonal part of the potential \eqref{Eq:WQ}, which breaks the $U(1)$ symmetry, and the diagonal part, which preserves the $U(1)$ symmetry. When $R \rightarrow 0$, the system approaches a protected metallic phase and Anderson's localization length diverges as $\Lambda(R) = \Lambda_0  \left ( \frac{R_0}{R}\right)^\nu$, where $\nu \simeq 2.6$ is the critical exponent of the metal-insulator transition for spin-Chern insulators \cite{HungPRB2016}. The index 0 indicates some reference configuration and, if $\Lambda_0$ and $R_0$ for this reference configuration are known, one can easily asses if Anderson's localization length is larger than the size of devices.
 
 \vspace{0.2cm}
 
 The last issue we want to investigate is how to define a topological bulk valley-Hall state in the presence of disorder. We proceed along the lines of \cite{ProdanPRB2009}. First, note that the generator of the $U(1)$ symmetry we ideally want to  preserve is:
 \begin{equation}
 \Xi = U \Sigma U^\dagger, \quad \Xi^\dagger=\Xi, \quad \Xi^2 = I.
 \end{equation}
We now introduce a weak bulk disorder:
\begin{equation}
{\bm H}_{\rm QVHE} \rightarrow {\bm H}_{\rm QVHE} + \begin{pmatrix} {\bm V}_\omega & 0 \\ 0 & {\bm V}_\omega \end{pmatrix},
\end{equation}
such that a spectral gap $G$ remains opened in the energy spectrum. Let $\bm P_G$ be its gap projection and consider the projected generator $\bm P_G \Xi \bm P_G$. If the disorder preserves the $U(1)$ symmetry, then ${\rm Spec}(\bm P_G \Xi \bm P_G)$ consists of $\pm 1$. If the $U(1)$ symmetry is slightly broken, the spectrum of $\bm P_G \Xi \bm P_G$ expands but a spectral gap remains, enough to define the valley-filtered projections $\bm P_G^{\pm}$ as the spectral projections onto the upper/lower spectrum of $\bm P_G \Xi \bm P_G$:
\begin{equation}
\bm P_{VG}^+ = \chi_{[0,1)}(\bm P_G \Xi \bm P_G), \quad \bm P_{VG}^+ = \chi_{[-1,0)}(\bm P_G \Xi \bm P_G).
\end{equation} 
In analogy with spin-Chern number \cite{ShengPRL2006,ProdanPRB2009}, we then define:
\begin{equation}
{\rm Ch}_V(\bm P_{G}) = \tfrac{1}{2}\big ({\rm Ch}(\bm P_{VG}^+) - {\rm Ch}(\bm P_{VG}^-) \big ).
\end{equation}
Adopting the terminology from \cite{ProdanPRB2009}, we call the spectral gap of $P_G \Xi P_G$ the valley gap. We now can formulate the following exact statement \cite{ProdanPRB2009}:

\vspace{0.2cm}

\noindent{\bf Statement:} ${\rm Ch}_V(P_G)$ remains quantized and non-fluctuating as long as the energy {\it and} valley gaps remain opened. The statement holds if the spectral gaps are replaced by mobility gaps.

\vspace{0.2cm}

Let us specify that there is no exact bulk-boundary correspondence for this bulk-topological invariant. However, for spin-Chern insulators, it is known that the spin edge-currents are not shut off by the disorder even when the $U(1)$ symmetry is broken \cite{BaldesCMP2013}. Furthermore, extensive numerical investigations have been shown that the spin-Chern number stabilizes a region of extended bulk energy spectrum, which carries the invariant, even in the presence of strong disorder \cite{ProdanJPA2011,XuPRB2012}. All these point to new effects that might be present in the context of valley-Hall physics and are worth investigating. Additionally, the new concept of valley gap can be another useful quantitative estimator for QVHE.

\subsection{Discussion of the QVHE regime}
\label{Sec:QVHERegime}

Based on our analysis, we can formulate a precise statement which identifies the conditions in which QVHE is certain to occur. Assume:
\begin{itemize}
\item A 2-dimensional translational and time-reversal invariant lattice system that displays two momentum-separated Dirac cones in the bulk spectrum, stabilized by a lattice symmetry such as the inversion in the case of graphene. This system serves as the reference system $H_0$. 
\item A finite-range translational and time-reversal invariant Hamiltonian $Q$, which breaks the stabilizing lattice symmetry, splits the Dirac cones and opens a bulk gap.
\item A finite-range Hamiltonian $\delta H$ whose hopping matrices coincide with those of $Q$ on one side of a DW and with $\mathcal I Q \mathcal I$ on the other side of the DW. 
\end{itemize}

\vspace{0.2cm}

\noindent {\bf Statement:} The deformation Hamiltonian $H(\lambda) = H_0 + \lambda \, \delta H$ necessarily crosses a QVHE regime when $\lambda$ is turned on. More precisely, as soon as $\lambda >0$, two counter-propagating chiral domain-wall modes will emerge crossing the bulk gap at the two Dirac nodes. The two modes are exponentially localized in space near the DW, though the characteristic localization length depends on the size of the bulk gap. The two modes are protected against a certain class of disorder perturbations, as well as slow spatially-varying disorder perturbations. Furthermore, for a generic disorder potential, the parameter $R$ in \eqref{Eq:R} can be used to asses the metallic character of the modes.

\vspace{0.2cm}

Our analysis also provides some very specific tips on how to optimize the effect and make it as robust possible. For this, the following conditions need to be satisfied:
\begin{itemize}
\item The Dirac cones need to be well separated in the momentum-space.
\item The valleys need to be sharp and deep when projected on $k_\|$.
\item The orientation of the DW needs to be as close as possible to the ideal one for which the line $(k_\|,k_\bot=0)$ intersects both valley points.
\item The transition across the DW needs to be as smooth as possible.
\item The parameter $\lambda$ should be chosen as large as possible but roughly not larger than the depth of the valleys, in order to enhance the localization of the modes along the DW.
\end{itemize}

Many of these aspects will appear and will be discussed again in our experimental investigation, which comes next.

\section{Coupled Spinners: A Versatile Platform}
\label{Sec:ExpPlatform}

In this section we describe in detail the reported experimental platform based on coupled spinners. The centers of the spinners are pinned, restricting them to a single rotational degree of freedom. Starting from such a basic mechanical system, it is possible to engineer mechanical building blocks with controlled number of degrees of freedom by simply staking two or more spinners on top of each other. Furthermore, the centers of the spinners can be pinned in any two-dimensional pattern and, due to the particular engineering of the degrees of freedom, the latter can be easily coupled in virtually any desirable way. The resulting experimental platform is extremely versatile and can be utilized to implement any quadratic 2-dimensional discrete Hamiltonian. Various ways of breaking the time-reversal symmetry will be discuss in our future works.

\begin{figure}
\includegraphics[width=0.9\linewidth]{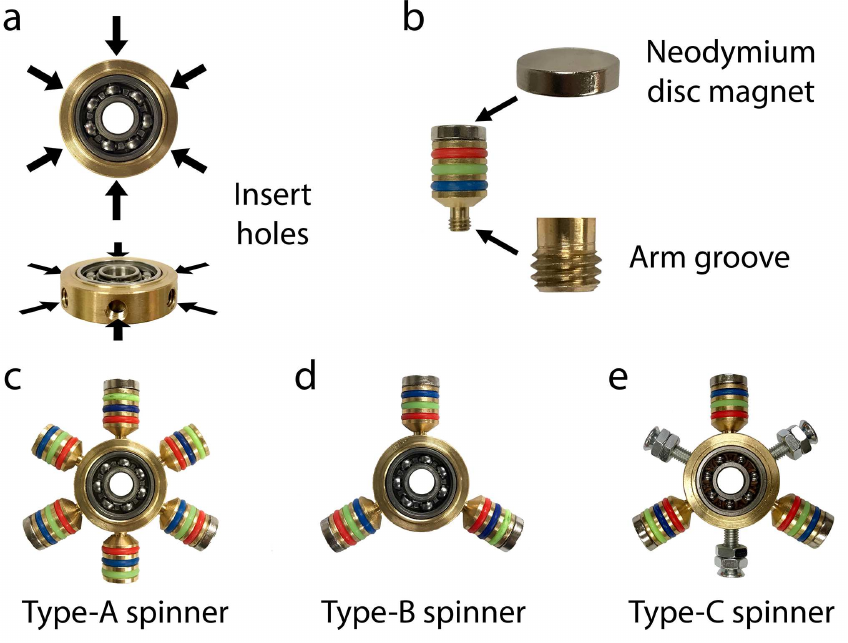}
\caption{\small {\bf Configurable spinner with detachable arms.} (a) Ball bearing with six inserts. (b) Detachable arms with magnetic ends for coupling. (c) Type-A spinner with connection arms and regular arms at 1, 3, 5 and 2, 4, 6 positions, respectively. (d) Type-B spinner with only connection arms at 1, 3, 5 positions. (e) Type-C spinner with connection arms and screws-and-nuts at 1, 3, 5 and 2, 4, 6 positions, respectively. }
\label{Fig:Spinner}
\end{figure}

\subsection{Configurable Spinners}

The configurable spinners are illustrated in Fig.~\ref{Fig:Spinner}. They consist of a stainless steel ball-bearing fixed in a brass encapsulation. The latter is fitted with six grooved indentations as shown in see Fig.~\ref{Fig:Spinner}(a), which enable us to attach additional components. This work features the relatively heavy brass arms shown Fig.~\ref{Fig:Spinner}(b), which can be securely fastened in the brass encapsulation via the end-bolts shown in the inset \ref{Fig:Spinner}(b). The arms are also fitted with strong magnets which provide the coupling between the spinners. This basic system enable us to achieve various configurations of the spinners as shown in \ref{Fig:Spinner}(c-e). Additional configurations can be achieved by modifying the brass encapsulations.  

\vspace{0.2cm}

The uniformity of the arms, their fastening to the encapsulation, the uniformity of the magnets, and the quality of the ball bearings are all essential for the proper functionality of the system. The latter reflects in the high Q-factors of the coupled resonators, which was measured to be around 50.

\begin{figure}
\center
  \includegraphics[]{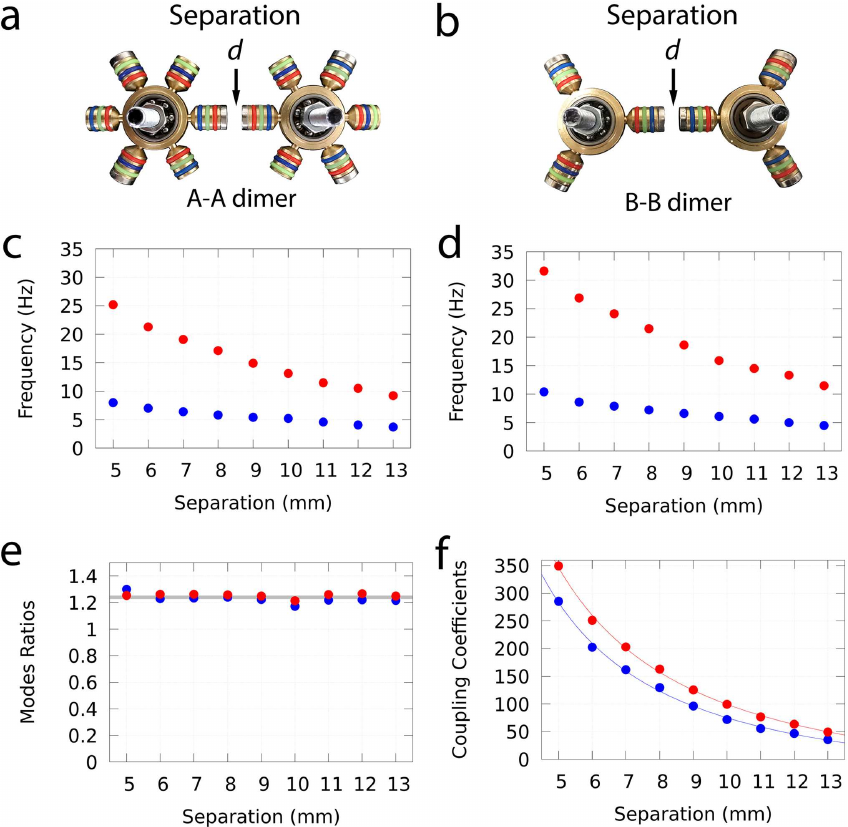}\\
  \caption{\small {\bf Mapping the coupling coefficients.} (a) The A-A dimmer configurations. (b) The experimentally measured resonant frequencies of the A-A dimer as functions of separation $d$ between the magnets. (c) The B-B dimmer configurations. (d) The experimentally measured resonant frequencies of the A-A dimer as functions of separation $d$ between the magnets. (e) The ratios $f_{\pm}^{A-A}/f_{\pm}^{B-B}$ as function of separation. (f) The coupling coefficients (solid dots) as derived from \eqref{Eq:CouplingCoeff} in units of $4\pi^2 I_A \times {\rm Hz}^2$, together with the analytic fits \eqref{Eq:Fit1} (continuous lines).}
 \label{Fig:Coupling1}
\end{figure}  

\subsection{Mapping the basic couplings}

The valley-Hall effect will be engineered with 6-arm and 3-arm spinners, called of Type-A and Type-B, respectively. As such, there are three basic couplings in our lattice: A-A, B-B and A-B. These magnetic couplings can measured by mapping the resonant modes of the corresponding dimers. The dynamics of a dimer is governed by the Lagrangian:
 \begin{equation}
L(\varphi_1,\varphi_2,\dot \varphi_1,\dot \varphi_2) = \tfrac{1}{2} I_1 \dot \varphi_1^2 + \tfrac{1}{2} I_2 \dot \varphi_2^2 - V(\varphi_1,\varphi_2).
\end{equation}
In the regime of small oscillations around the equilibrium configuration $\varphi_1=\varphi_2=0$, the potential can be approximated quadratically:
\begin{equation}\label{Eq:QuadraticV}
V(\varphi_1,\varphi_2) = V_0 + \tfrac{1}{2} \alpha (\varphi_1^2 + \varphi_2^2) + \beta \varphi_1 \varphi_2.
\end{equation}
The symmetry of the potential with respect to the exchange $1 \leftrightarrow 2$ is made explicit in this expansion. We will refer to $\alpha$ and $\beta$ as the coupling coefficients. The equations of motion are straightforward:
\begin{equation}
I_j \ddot \varphi_j + \alpha \varphi_j + \beta \varphi_{j'}=0, \quad j=1,2, \quad j'=2,1.
\end{equation}
With the ansatz $\varphi_j(t) = \frac{1}{\sqrt{I_j}}A_j e^{ i \omega t}$, $\omega=2\pi f$, the equation for the resonant modes reads:
\begin{equation}
\omega^2  \begin{pmatrix} A_1 \\ A_2 \end{pmatrix} = \begin{pmatrix} \tfrac{\alpha}{I_1} & \tfrac{\beta}{\sqrt{I_1 I_2}} \\ \tfrac{\beta}{\sqrt{I_1I_2}} & \tfrac{\alpha}{I_2} \end{pmatrix} \begin{pmatrix} A_1 \\ A_2 \end{pmatrix}.
\end{equation}
For $I_1 = I_2 = I_A$ or $I_B$, it leads to the pairs of resonant frequencies:
\begin{equation}\label{Eq:ResFreq1}
f_{\pm}^{A-A} = \sqrt{\frac{\alpha \pm \beta}{4 \pi^2 I_A}} \ , \quad f_{\pm}^{B-B} = \sqrt{\frac{\alpha \pm \beta}{4 \pi^2 I_B}}.
\end{equation}
The upper/lower frequency modes correspond to motions where the two angles are locked at $\varphi_2 = \pm \varphi_1$, respectively.

\begin{figure}
\center
  \includegraphics[]{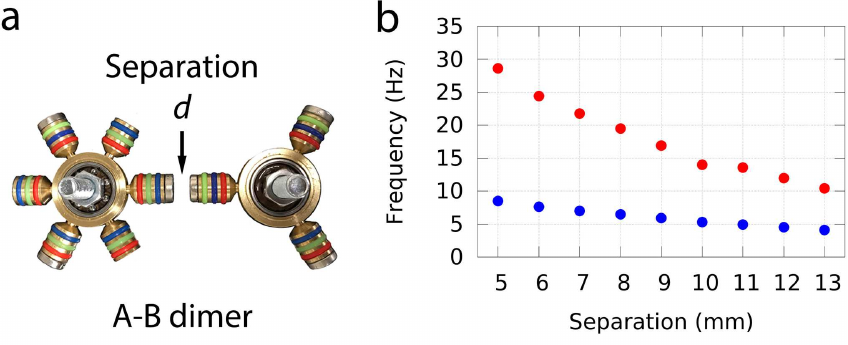}\\
  \caption{\small {\bf The A-B coupling.} (a) Experimental setup for the A-B dimer. (b) Experimentally measured resonant frequencies of the A-B dimer.}
 \label{Fig:Coupling2}
\end{figure}

\vspace{0.2cm}

The resonant frequencies have been independently measured as functions of distance, $d$, between the magnets and the data is reported in Fig.~\ref{Fig:Coupling1}(c, d). At this point we can verify if the coupling coefficients are affect by the removal of the arms, by examining the ratios $f_{\pm}^{B-B}/f_{\pm}^{A-A}$. As one can see in Fig.~\ref{Fig:Coupling1}(e), these two ratios are more or less identical and independent of $d$. From \eqref{Eq:ResFreq1}, this constant and common value can be identified with the ratio $\sqrt{\frac{I_A}{I_B}}$, which comes to $1.235$ from the fit, hence we obtain a quantitative measure of $r=I_A/I_B = 1.525$. Furthermore, it is possible to invert any of the relations in \eqref{Eq:ResFreq1}, {\it e.g.} A-A, and map the coupling coefficients:

\begin{equation}\label{Eq:CouplingCoeff}
\alpha = 2 \pi^2 I_A \big (f_+^2 + f_-^2 \big ), \quad \beta = 2 \pi^2 I_A \big (f_+^2 - f_-^2 \big ).
\end{equation}

The resulting values are shown in Fig.~\ref{Fig:Coupling1}(f), in convenient units of $4\pi^2 I_A  \times {\rm Hz}^2$, together with the theoretical fits:


\begin{align}\label{Eq:Fit1}
\begin{matrix}
\alpha(d)=-\tfrac{654.09}{\sqrt{d}} + \tfrac{2763.66}{d}+\tfrac{575.89}{d^2},\\
\beta(d)=-\tfrac{778.14}{\sqrt{d}} + \tfrac{3439.81}{d}+\tfrac{161.35}{d^2}.
\end{matrix}
\end{align}

For completeness, the resonant frequencies for the A-B dimer are reported in Fig.~\ref{Fig:Coupling2}. They agree well with the coupling coefficients \eqref{Eq:Fit1}.

\begin{figure}
\center
  \includegraphics[width=\linewidth]{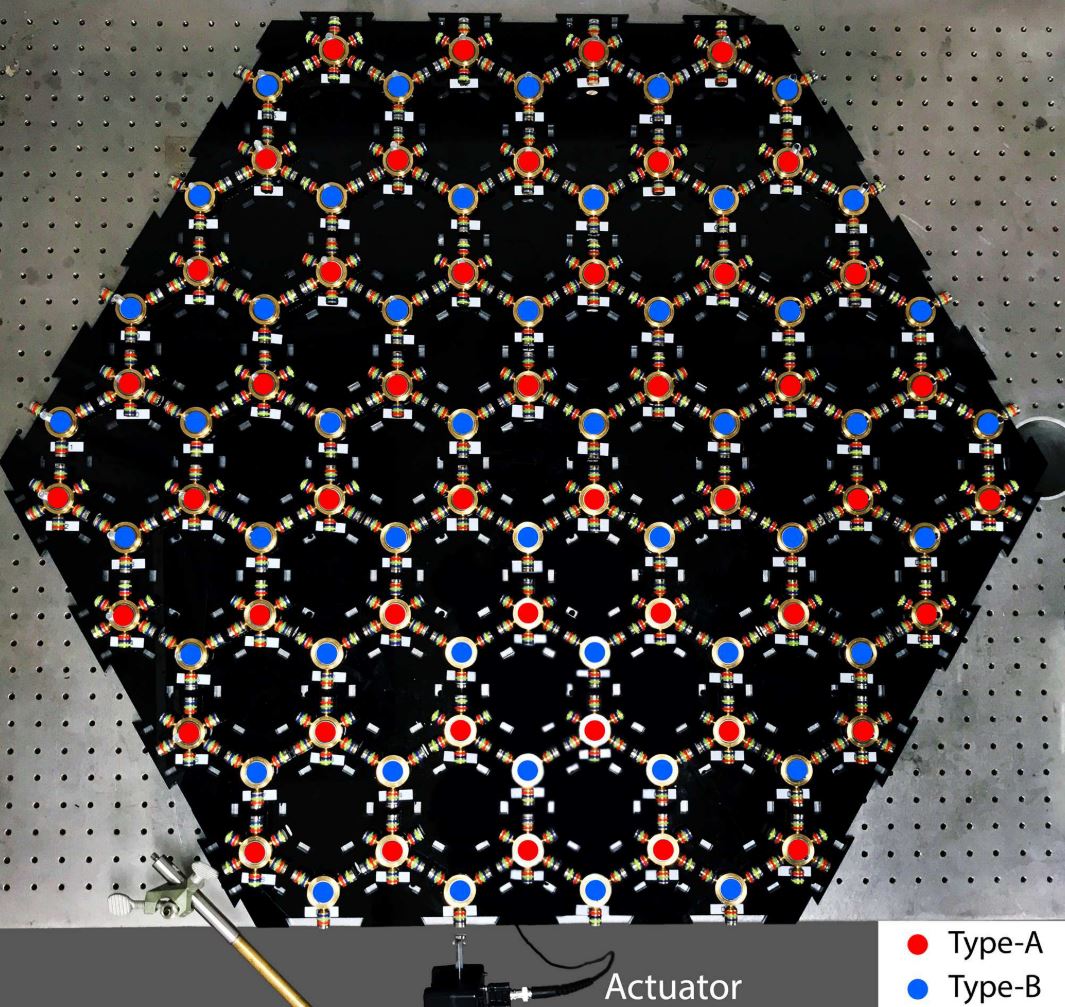}\\
  \caption{\small {\bf The bulk configuration of spinners.} It is a finite bipartite honeycomb lattice populated with A (red) and B (blue) type spinners. The actuator appears at the bottom of the illustration. For technical specifications, see section~\ref{Sec:ExpTechnique}.}
 \label{Fig:BulkSystem}
\end{figure}

\section{Bulk Analysis}
\label{Sec:Bulk}

\begin{figure}
\center
  \includegraphics[width=0.6\linewidth]{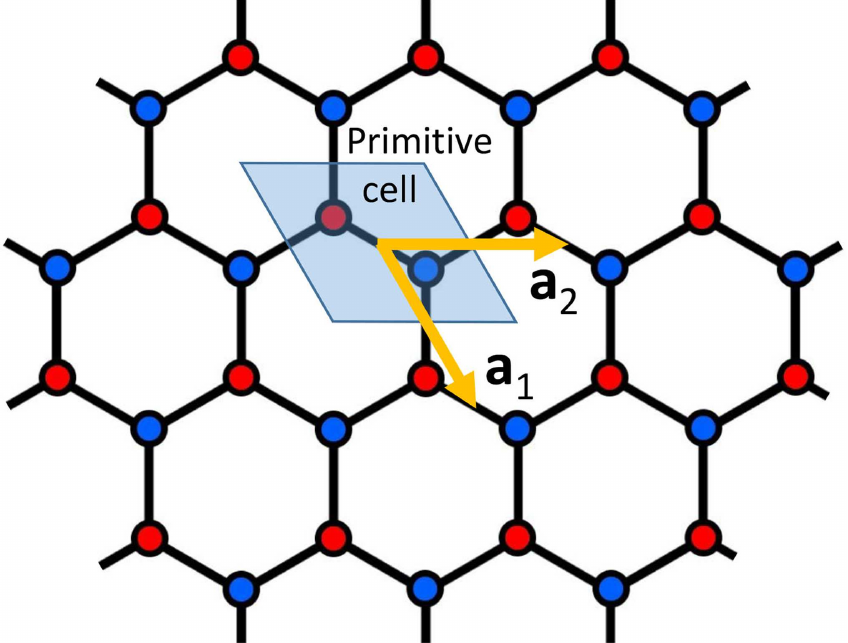}\\
  \caption{\small {\bf Honeycomb lattice.} The primitive cell is indicated by the shaded area and the primitive vectors $\bm a_{1,2}$ are indicated by the arrows.}
 \label{Fig:Lattice}
\end{figure}

The fully assembled spinner system is shown in Fig.~\ref{Fig:BulkSystem}. The spinners are arranged in a honeycomb lattice whose primitive cell and primitive vectors are shown in Fig.~\ref{Fig:Lattice}.

\subsection{Mapping the Spectrum}

 The centers of the primitive cells are located at $\bm R_{\bm n} = n_1 \bm a_1 + n_2 \bm a_2$, hence we can label the cells by $\bm n = (n_1,n_2) \in \mathbb Z^2$. The spinners can be easily identified using the pair of indexes $(\bm n, A)$ or $(\bm n, B)$. It is convenient to introduce the shift operations on the indices:
 
\begin{align}\label{Eq:Shifts}
\begin{matrix}
S_1 \bm n = S_1 (n_1,n_2) = (n_1+1,n_2),\\
S_2 \bm n = S_2 (n_1,n_2) = (n_1,n_2+1).
\end{matrix}
\end{align}
With these notations, the Lagrangian of the infinite lattice takes the form:
\begin{align}
\mathcal L = \sum_{\bm n \in \mathbb Z^2} \big (&\tfrac{1}{2} I_A \dot \varphi_{\bm n,A}^2 + \tfrac{1}{2} I_B \dot \varphi_{\bm n,B}^2 - V(\varphi_{\bm n,A},\varphi_{\bm n,B}) \\
\nonumber & - V(\varphi_{\bm n,A},\varphi_{S_1^{-1}\bm n,B}) - V(\varphi_{\bm n,A},\varphi_{S_2^{-1}\bm n,B} \big ).
\end{align}
Using the quadratic approximation \eqref{Eq:QuadraticV}, the equations of motions take the form:
\begin{align} \label{Eq:Dyn1}
\begin{pmatrix} I_A \ddot \varphi_{\bm n,A} \\ I_B \ddot \varphi_{\bm n,B} \end{pmatrix}=\begin{pmatrix}-3\alpha \varphi_{\bm n,A} - \beta( \varphi_{\bm n,B}+\varphi_{S_1^{-1}\bm n,B}+\varphi_{S_2^{-1}\bm n,B}) \\-3\alpha \varphi_{\bm n,B} - \beta( \varphi_{\bm n,A}+\varphi_{S_1\bm n,A}+\varphi_{S_2 \bm n,A})\end{pmatrix}
\end{align}
It is convenient to make the change of variables:
\begin{equation}
\varphi_{\bm n,A} = \frac{1}{\sqrt{I_A}} \psi_{\bm n,A}, \quad \varphi_{\bm n,B} = \frac{1}{\sqrt{I_B}} \psi_{\bm n,B},
\end{equation}
and bring the equations to the form:

\begin{align}
\begin{pmatrix}
\ddot \psi_{\bm n,A} \\
\ddot \psi_{\bm n,B}\end{pmatrix}=
\begin{pmatrix}
-3\tfrac{\alpha}{I_A} \psi_{\bm n,A} - \tfrac{\beta}{\sqrt{I_A I_B}}( \psi_{\bm n,B}+\psi_{S_1^{-1}\bm n,B}+\psi_{S_2^{-1}\bm n,B})\\
-3\tfrac{\alpha}{I_B} \psi_{\bm n,B} - \tfrac{\beta}{\sqrt{I_A I_B}} ( \psi_{\bm n,A}+\psi_{S_1\bm n,A}+\psi_{S_2\bm n,A})
\end{pmatrix}
\end{align}
We can encode the degrees of freedom in a single function:
\begin{equation}
\bm \psi : \mathbb Z^2 \rightarrow \mathbb C^2, \quad \bm \psi(\bm n)={\small \begin{pmatrix} \psi_{\bm n,A} \\ \psi_{\bm n,B}\end{pmatrix}}
\end{equation}
and use the ansatz $\bm \psi(t) \rightarrow {\rm Re}\big [e^{i \omega t} \bm \psi \big ]$, $\omega = 2\pi f$. Then, in the units used in Fig.~\ref{Fig:Coupling1}, the equations of motions simplify to:
\begin{align}
\begin{matrix}
f^2 \bm \psi = \bigg [ \tfrac{3}{2}\alpha\big (1+r + (1-r) \sigma_3\big )  \\
+ \beta \sqrt{r} \big (\sigma_1 + \sigma_- (S_1+S_2)+\sigma_+  (S_1^\dagger  + S_2^\dagger) \big ) \bigg ] \bm \psi,
\end{matrix}
\end{align}
where the shift operators act as:
\begin{equation}
(S_j \bm \psi)(\bm n) = \bm \psi(S_j \bm n), \quad j=1,2,
\end{equation} 
and $\sigma$'s are Pauli's matrices. The shift operators commute with each other and with the dynamical matrix, and have common eigenvectors:
\begin{equation}
S_j \, e^{i{\bm k} {\bm n}} = e^{i k_j} e^{i{\bm k} {\bm n}}, \quad \bm k \in [-\pi,\pi]^2, \quad j=1,2. 
\end{equation}
Hence, the normal modes will be sought in the form $\bm \psi(\bm n)=e^{i{\bm k} {\bm n}}\, \xi$, $\xi \in \mathbb C^2$, in which case the dispersion equation reduces further to:
\begin{align}
f^2 \xi = \bigg (\tfrac{3}{2}\alpha (1+r +(1-r) \sigma_3)  + \beta \sqrt{r} {\small \begin{pmatrix} 0 & \gamma(\bm k)^\ast \\ \gamma(\bm k) & 0 \end{pmatrix}} \bigg ) \xi,
\end{align}
with $\gamma(\bm k) = 1 + e^{i k_1} + e^{i k_2}$. The explicit dispersion equations of the resonant modes then follow:
\begin{equation}\label{Eq:BulkDispersion}
f_\pm = \bigg [\tfrac{3\alpha}{2}(1+r) \pm \sqrt{\tfrac{9\alpha^2}{4}(1-r)^2 +r \beta^2|\gamma(\bm k)|^2} \, \bigg ]^\frac{1}{2}.
\end{equation}
When $r=1$, the system is inversion symmetric and two Dirac cones are present. The imbalance between $I_A$ and $I_B$ breaks the inversion symmetry, hence the Dirac cones split as soon as $r>1$. Let us note that the valleys are located at the points where $|\gamma(\bm k)|=0$, which are $K=-K'=(\frac{2\pi}{3},-\frac{2\pi}{3})$.

\begin{figure}
\center
  \includegraphics[]{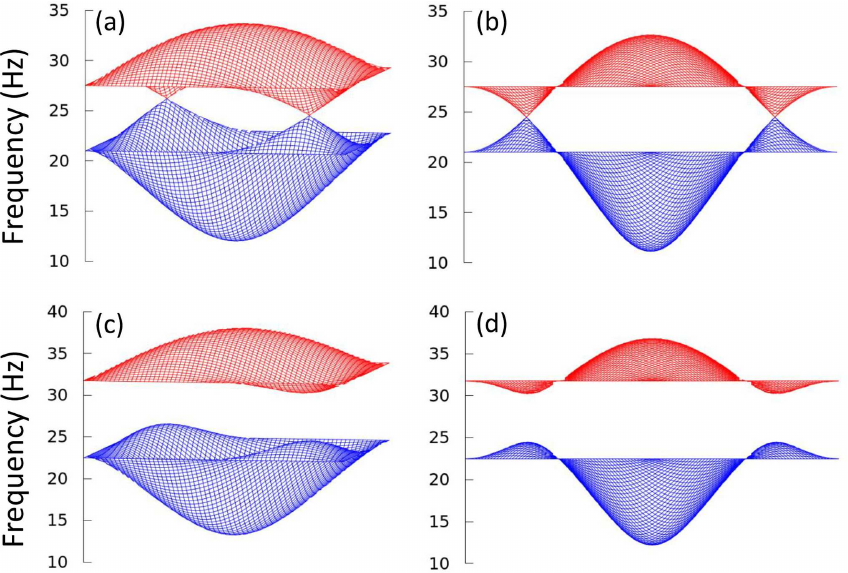}\\
  \caption{\small {\bf Theoretical bulk spectrum.} (a,b) $I_A=I_B$ and (c,d) $I_A/I_B=1.525$ (the experimental value). The left columns give a 3-dimensional view while the right columns show a projection on the plane determined by the second diagonal of the $(k_1,k_2)$ plane.}
 \label{Fig:TheoreticalBulk}
\end{figure}

\vspace{0.2cm}

\begin{figure}
\center
  \includegraphics[]{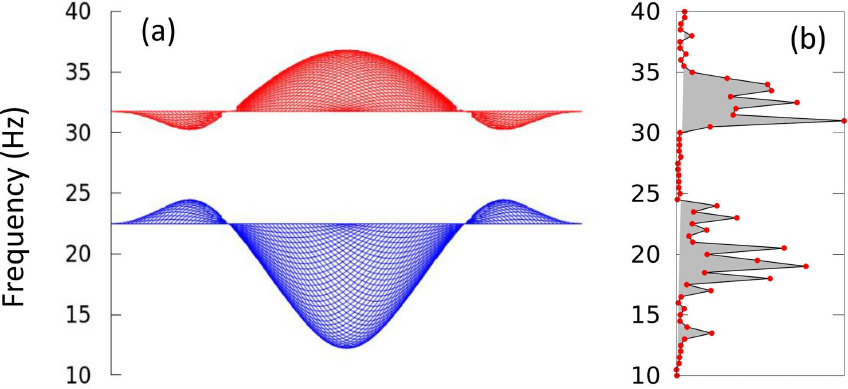}\\
  \caption{\small {\bf Theoretical versus experimental bulk spectrum}. (a) The theoretical data taken from panel (d) of Fig.~\ref{Fig:TheoreticalBulk}. (b) The reading from a sensor placed inside the spinner structure when the system is actuated from the edge as seen in Fig.~\ref{Fig:BulkSystem}.}
 \label{Fig:BulkThVsExp}
\end{figure}

A graphical representation of the dispersions \eqref{Eq:BulkDispersion} is reported in Fig.~\ref{Fig:TheoreticalBulk}, for both $r=1$ and the experimental value $r=1.525$. A comparison between the theoretical spectrum and the experimental reading from a sensor placed inside the structure is shown in Fig.~\ref{Fig:BulkThVsExp}. The details of the measurements are reported in section \ref{Sec:ExpTechnique}. As one can see, the agreement in Fig.~\ref{Fig:BulkThVsExp} is excellent.

\begin{figure}[b]
\center
  \includegraphics[width=1\linewidth]{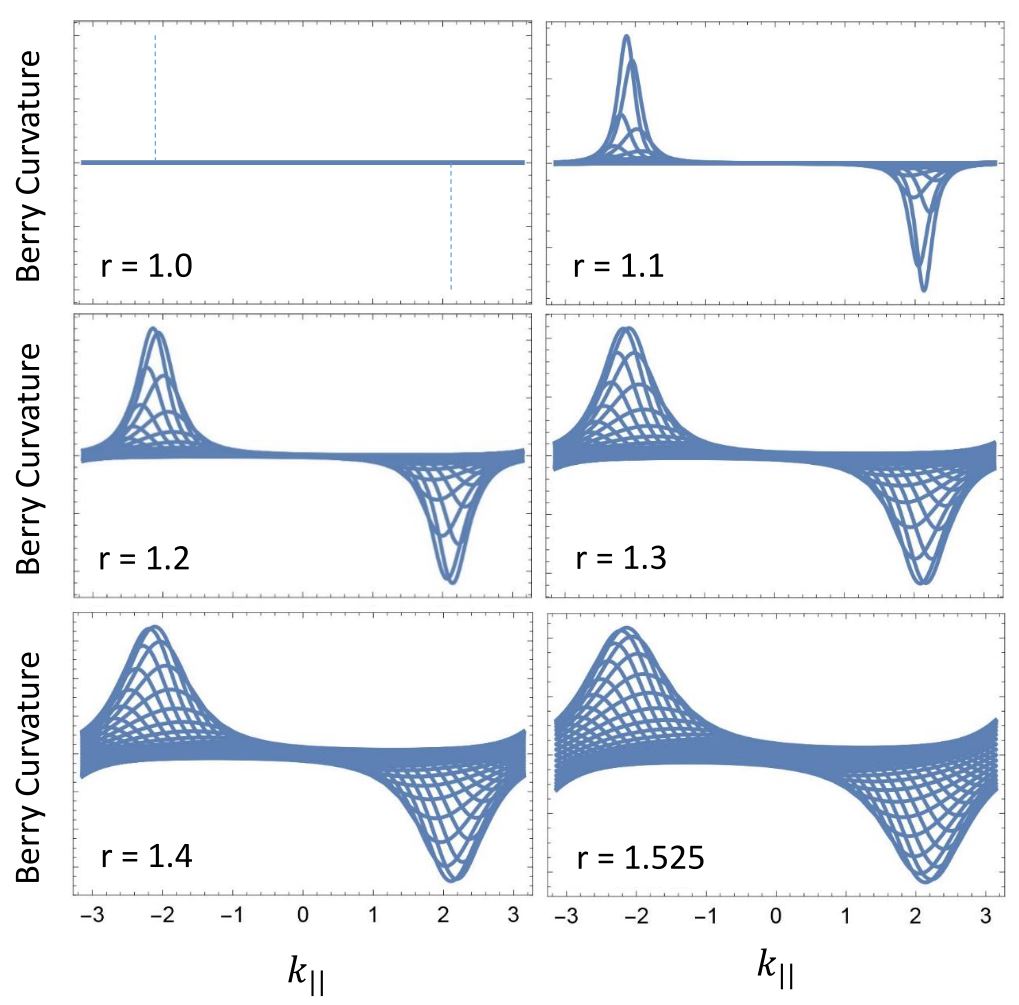}\\
  \caption{\small {\bf Berry curvature.} The theoretical calculations were performed with the experimental values of the coupling constants $\alpha$ and $\beta$ and for the specified values of $r=I_A/I_B$. The last value $r=1.525$ is the experimental value. Only the relevant values along the second diagonal of the $\bm k$-plane are shown. The data is rendered as function of $k_\|=\frac{1}{2}(k_1-k_2)$.}
 \label{Fig:BerryCurvature}
\end{figure}

\subsection{Mapping the Berry Curvature}

Here we map the Berry curvature \eqref{Eq:Curvature}. For a two-band model, the gap projector can be expressed as:
\begin{equation}
P_G (\bm k) = \tfrac{1}{2}(I - {\bm n} \cdot {\bm \sigma}) ,
\end{equation} 
where $ {\bm n}(\bm k) = (n_{x},n_{y},n_{z})$ is the unit vector on the Bloch sphere, which in our case is given by: 
\begin{equation}
{\bm n}(\bm k) = \frac{\bigg( \sqrt{r} \beta {\rm Re}[ \gamma(\bm k) ], \sqrt{r} \beta {\rm Im}[ \gamma(\bm k) ], \tfrac{3\alpha}{2}(1 - r) \bigg)}{ \sqrt{\tfrac{9\alpha^2}{4}(1-r)^2 +r \beta^2|\gamma(\bm k)|^2})}.
\end{equation}
Then $F(\bm k) $ reduces to:
\begin{equation} 
F(\bm k) = \pi \,  {\bm n} \cdot (\partial_{k_{1}} {\bm n} \times \partial_{k_{2}} {\bm n}),
\end{equation}
which leads to the explicit expression:
\begin{equation}
F(\bm k) = \frac{\pi}{2} \frac{3(1-r)r \alpha \beta^2 \sin(k_1-k_2)}{ \bigg(\frac{9 \alpha^2}{4} (1-r)^2+ r \beta^2 |\gamma(\bm k)|^2 \bigg)^{3/2}}.
\end{equation}
A graphical representation is given in Fig.~\ref{Fig:BerryCurvature} for increasing values of $r$. Practically, for all those values, the conditions stated in section~\ref{Sec:QVHERegime} regarding the Berry curvature are met, provided we take the domain wall in the direction of $\bm a_1 - \bm a_2$.

\begin{figure}
\center
\includegraphics[width=\linewidth]{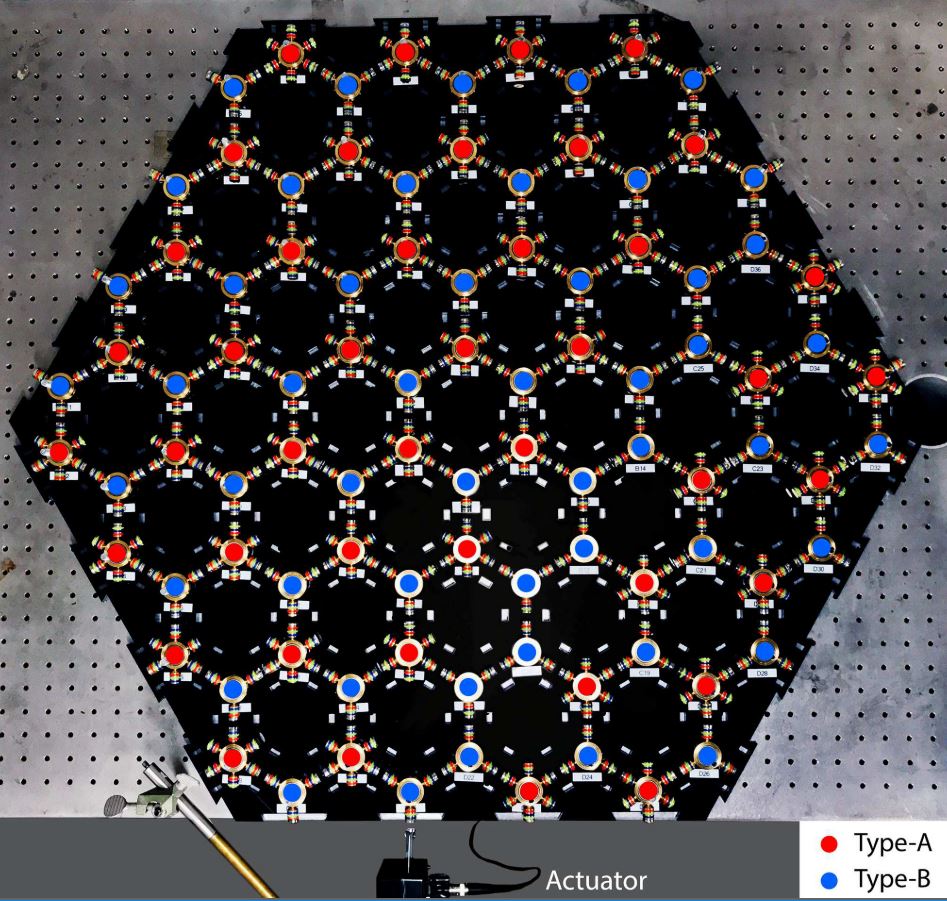}\\
\caption{\small {\bf The experimental setup with a straight domain-wall.} The domain wall consists of the zig-zag chain of B type (blue) spinners. Note the actuator positioned at one end of the domain-wall.}
 \label{Fig:ExpStraightWall}
\end{figure}

\section{Domain-Wall Analysis}
\label{Sec:DomWall}

\begin{figure}
\center
  \includegraphics[width=0.8\linewidth]{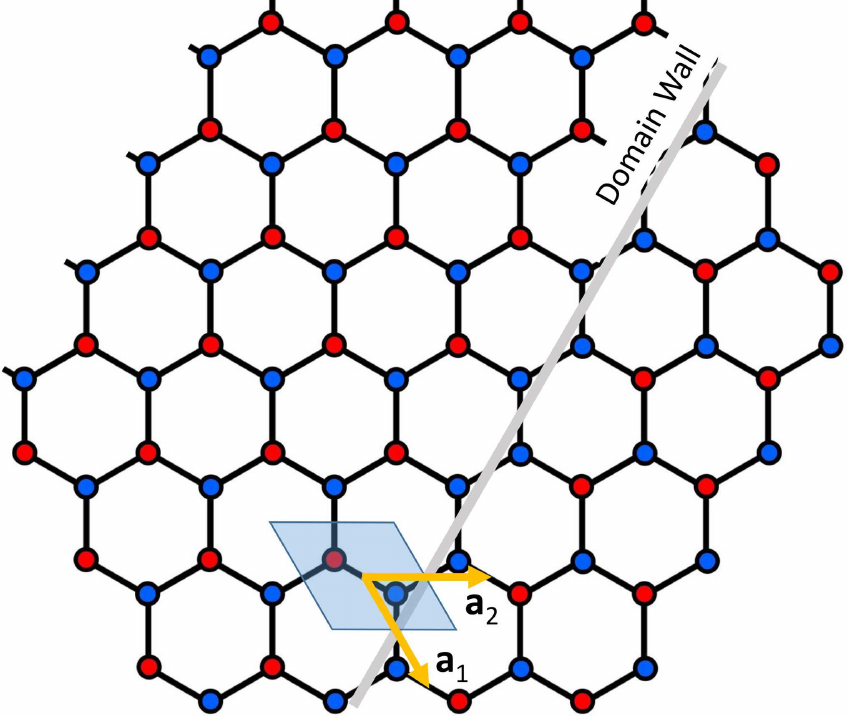}\\
  \caption{\small {\bf Schematic of the domain-wall.} Note that the domain-wall does not cut any of the primitive cells.}
 \label{Fig:SchematicStraightWall}
\end{figure}

The experimental setup with a straight DW is shown in Fig.~\ref{Fig:ExpStraightWall}. A schematic and more geometrical data are shown in Fig.~\ref{Fig:SchematicStraightWall}. It is important to note that the domain-wall is consistent with the previously chosen primitive cell.

\subsection{Theoretical Considerations}
\label{Sec:TC}

Since the domain wall is along $\bm a_\|=\bm a_1-\bm a_2$, we will switch to the shift operators  $S_2$ and:
\begin{equation}
S_\|=S_{\bm a_1 - \bm a_2} = S_1 S_2^\dagger \ \ \Rightarrow \ \ S_1 =   S_\| S_2.
\end{equation}
Furthermore, since the positions of the centers of the primitive cells can be expressed as:
\begin{equation}
\bm R_{\bm n} = n_1 \bm a_1+n_2 \bm a_2 = n_1 \bm a_\| + (n_1+n_2) \bm a_2,
\end{equation} 
we relabel the sites by $(m,n)=(n_1,n_1+n_2)$. Note that the shift operators now act as:
\begin{equation}
S_\| |m,n\rangle = |m+1,n\rangle, \quad S_2 |m,n\rangle = |m,n+1 \rangle,
\end{equation}
and the primitive cells just below the domain-wall all have $n=0$ and the ones above the domain-wall have $n=-1$. Furthermore, the reflection $\mathcal I$ acts as: $\mathcal I |m,n,A\rangle = |m,-n-1,B\rangle$ and $\mathcal I |m,n,B\rangle = |m,-n-1,A\rangle$.

\vspace{0.2cm}

To determine the dispersion equation, we go back to Eq.~\ref{Eq:Dyn1}. Note that, as we cross the domain wall, the coupling coefficients $\alpha$ and $\beta$ remain the same and the only effect is the exchange of $I_A$ and $I_B$. Hence, in the presence of the domain wall, these equations become:
\begin{align}\label{Eq:Partial1}
\omega^2 \Big(I - \delta I \, \sigma_3 \,{\rm sgn}(X_2)\Big) \bm \varphi =  D \bm \varphi,
\end{align} 
where $I=\frac{1}{2}(I_A+I_B)$ and $\delta I = \frac{1}{2}(I_A-I_B)$, as well as:
\begin{equation}
{\rm sgn}(X_2) = \sum_{n,m} {\rm sgn}(n) \, |m,n \rangle \langle m,n|,
\end{equation}
with the convention that ${\rm sgn}(0) = 1$. The dynamical matrix given by:
\begin{equation}\label{Eq:DynM}
D = 3\alpha+\beta \big (\sigma_1 + \sigma_- (S_\| S_2+S_2)+\sigma_+  (S_2^\dagger S_\|^\dagger  + S_2^\dagger) \big ).
\end{equation}
Note that $\delta I=0$ represents the reference system in our theoretical analysis. We transform \eqref{Eq:Partial1} in a standard eigen-system problem by performing the transformation:
\begin{equation}
\bm \varphi = \big (I - \delta I \, \sigma_3 \, {\rm sgn}(X_2)\big)^{-\frac{1}{2}} \bm \psi,
\end{equation}
in which case the dispersion equation becomes:
\begin{equation}
\omega^2 \bm \psi = \sqrt{I - \delta I \, \sigma_3 \, {\rm sgn}(X_2)} \, D \, \sqrt{I - \delta I \, \sigma_3 \, {\rm sgn}(X_2)} \bm \psi.
\end{equation}
Note that, with $\Gamma = \frac{1}{2}\big (\frac{1}{\sqrt{I_A}}+\frac{1}{\sqrt{I_B}} \Big )$ and $\Delta = \frac{1}{2}\big (\frac{1}{\sqrt{I_A}}-\frac{1}{\sqrt{I_B}} \Big )$:
\begin{equation}
\sqrt{I - \delta I \, \sigma_3 \, {\rm sgn}(X_2)}= \Gamma -\Delta \, \sigma_3 \, {\rm sgn}(X_2).
\end{equation}

\vspace{0.2cm}

Lastly, because $S_\|$ commutes with the dynamical matrix, we can seek the modes in the form:
\begin{equation}
\bm \psi(m,n) = {\rm Re}\big [ e^{ik m} \bm \xi_k(n) \big ], \quad k \in [-\pi, \pi],
\end{equation}
in which case the dispersion equation takes the form $H(k)\bm \xi_k = \omega_k^2 \bm \xi_k$, where $H$ relates to the operator in \eqref{Eq:OrigHam} and has the explicit expression:
\begin{equation}\label{Eq:EdgeDispersion}
H(k) = \big (\Gamma - \Delta \sigma_3 {\rm sgn}(X_2)\big) \, D_k \, \big (\Gamma - \Delta \sigma_3 {\rm sgn}(X_2)\big), 
\end{equation}
with $D_k$ derived directly from \eqref{Eq:DynM}:
\begin{equation}
D_k = 3\alpha+\beta\big (\sigma_1 + (e^{ik}+1)\sigma_-  S_2+(e^{-ik}  + 1) \sigma_+  S_2^\dagger \big ).
\end{equation}
Note that $k$ stands here for $k_\|$ and the relation between $k$ and $(k_1,k_2)$ used in Fig.~\ref{Fig:BulkThVsExp} is $k=\frac{1}{2}(k_1-k_2)$.
The action of $H(k)$ on the basis of $\CM^2 \otimes \ell^2(\ZM)$:
\begin{equation}
|n,+1 \rangle = {\small\begin{pmatrix} |n \rangle \\ 0 \end{pmatrix}}, \quad |n,-1\rangle = {\small\begin{pmatrix} 0 \\ |n \rangle \end{pmatrix}},
\end{equation}
can be written explicitly ($s = \pm 1$):
\begin{align}
\begin{matrix}
H(k) \, |n,s\rangle = 3\alpha \big (\Gamma - s \Delta \, {\rm sign}(n) \big)^2 |n,s\rangle \\
+ \beta (\Gamma^2 - \Delta^2) |n,-s\rangle 
 + \beta (\Gamma^2 - \Delta^2)(e^{isk}+1) |n+s,-s\rangle.
 \end{matrix}
\end{align}
With this expression, one can convince himself that $H$ can be indeed written in the form $H_0+\lambda \, \delta H$ as in section~\ref{Sec:QVHERegime}, with $\lambda=1-r$. The predictions can now be checked both numerically and experimentally.

\begin{figure}
\center
  \includegraphics[width=\linewidth]{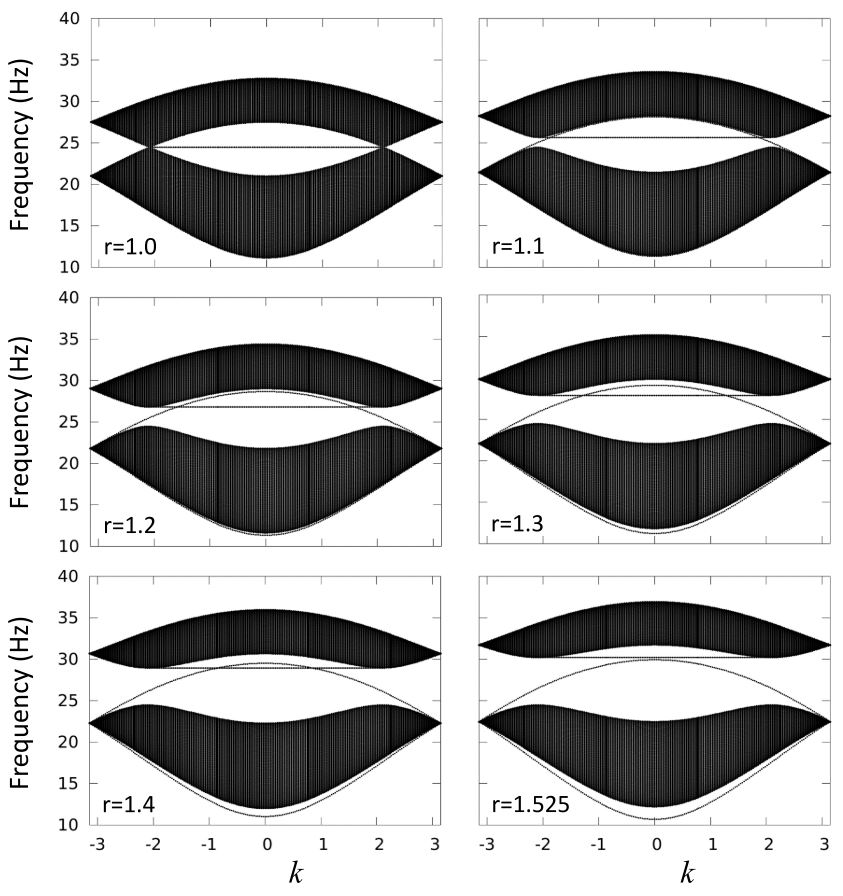}\\
  \caption{\small {\bf Theoretical spectrum in the presence of a domain-wall.} Simulations are shown for increasing values of $r=\frac{I_A}{I_B}$, ranging from 1 all the way to the experimental value of 1.525. The spectrum is computed on a strip with the domain-wall at the center. The doubly degenerated flat band seen in all panels is located at the edges of the ribbon, hence it is un-related to the physics studied here.}
 \label{Fig:TheoreticalEdge}
\end{figure}

\vspace{0.2cm}

The resulting spectrum for the domain-wall configuration of Fig.~\ref{Fig:ExpStraightWall} is reported in Fig.~\ref{Fig:TheoreticalEdge}. It reproduces the well-known QVHE features and fully confirms our theoretical predictions. More precisely, for small values if $r-1$, the effect is strong and predictable but, as this parameter is increased, one can see that the effect will eventually disappear. This exemplifies the fact that QVHE is a transient effect, {\it i.e.} it can vanish without a phase transition, but nevertheless there is a predictable region of the parameter space where the effect will always be strong and reliable.

\subsection{Experimental Observation of the DW Modes}

The DW has been actuated from one end, as shown in Fig.~\ref{Fig:ExpStraightWall}, until a stationary regime was established. In this setup, the counter-propagating DW modes are scattered into each other at the ends of the interface, leading to a standing wave. Pick-up coils similar to the ones found in electric guitars have been placed on top of the bonds and the standing wave pattern was mapped out. While the details of the measurements are provided in the following section, let us mentioned that four magnetic bonds in the $\bm a_2$ direction have been probed, starting from the DW, enough to asses the spatial localization of the modes.

\begin{figure*}
\center
  \includegraphics[width=0.8\textwidth]{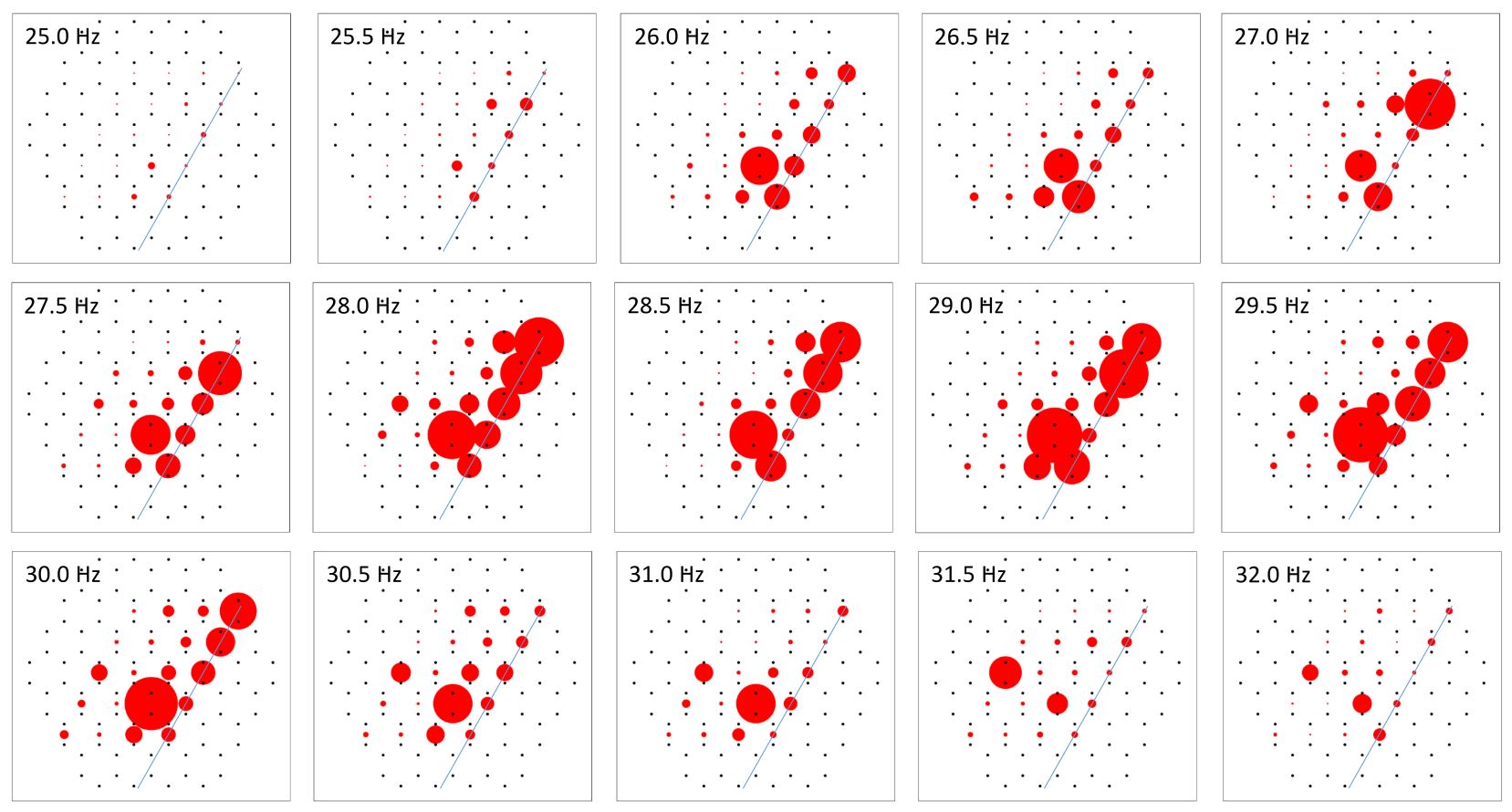}\\
  \caption{\small {\bf Experimental observation of the DW modes.} The fine line marks the position of the interface relative to the honeycomb lattice, indicated by the black dots. The red disks mark the position of the motion sensors, which are placed above bonds. The size of a disk is proportional with the reading of the motion sensor at that location. The frequencies, which are marked in each panel, sample the entire bulk gap.}
 \label{Fig:DataSEdge}
\end{figure*}

\begin{figure}[b]
\center
\includegraphics[width=\linewidth]{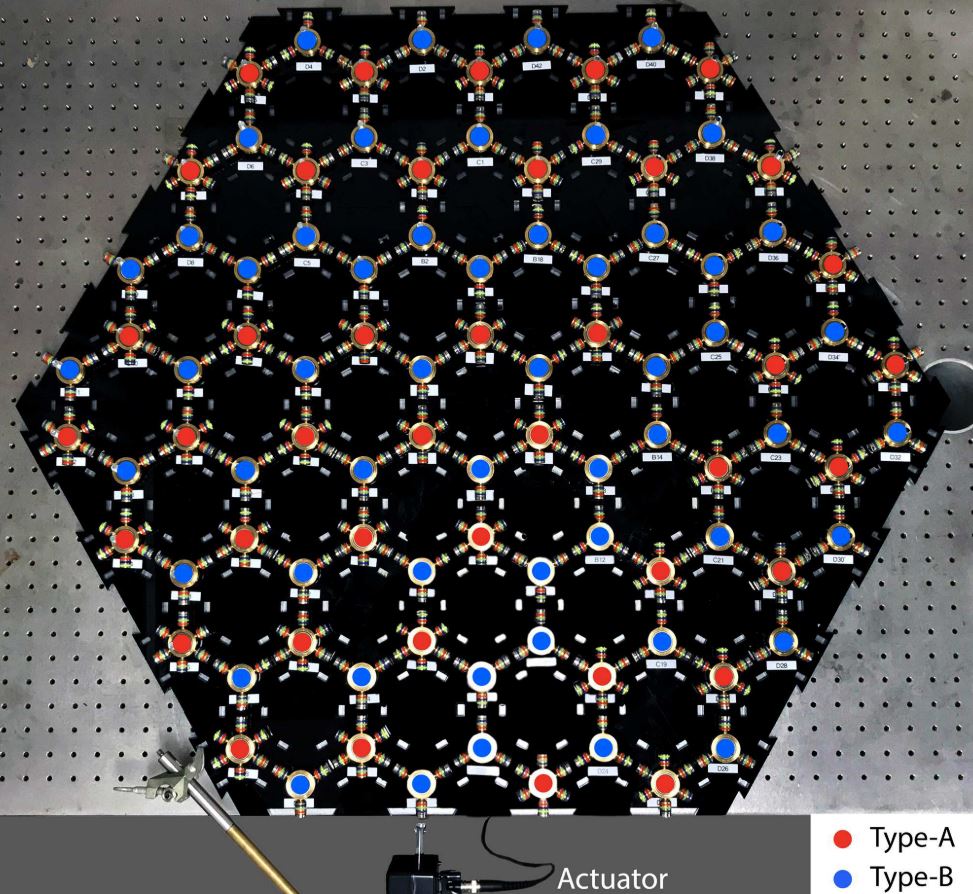}\\
\caption{\small {\bf The experimental setup with a L-shape domain-wall.} The domain wall consists of the zig-zag chain of B type (blue) spinners. Note again the actuator positioned at one end of the domain-wall.}
 \label{Fig:ExpLShapeWall}
\end{figure}

\begin{figure*}[t]
\center
  \includegraphics[width=0.8\textwidth]{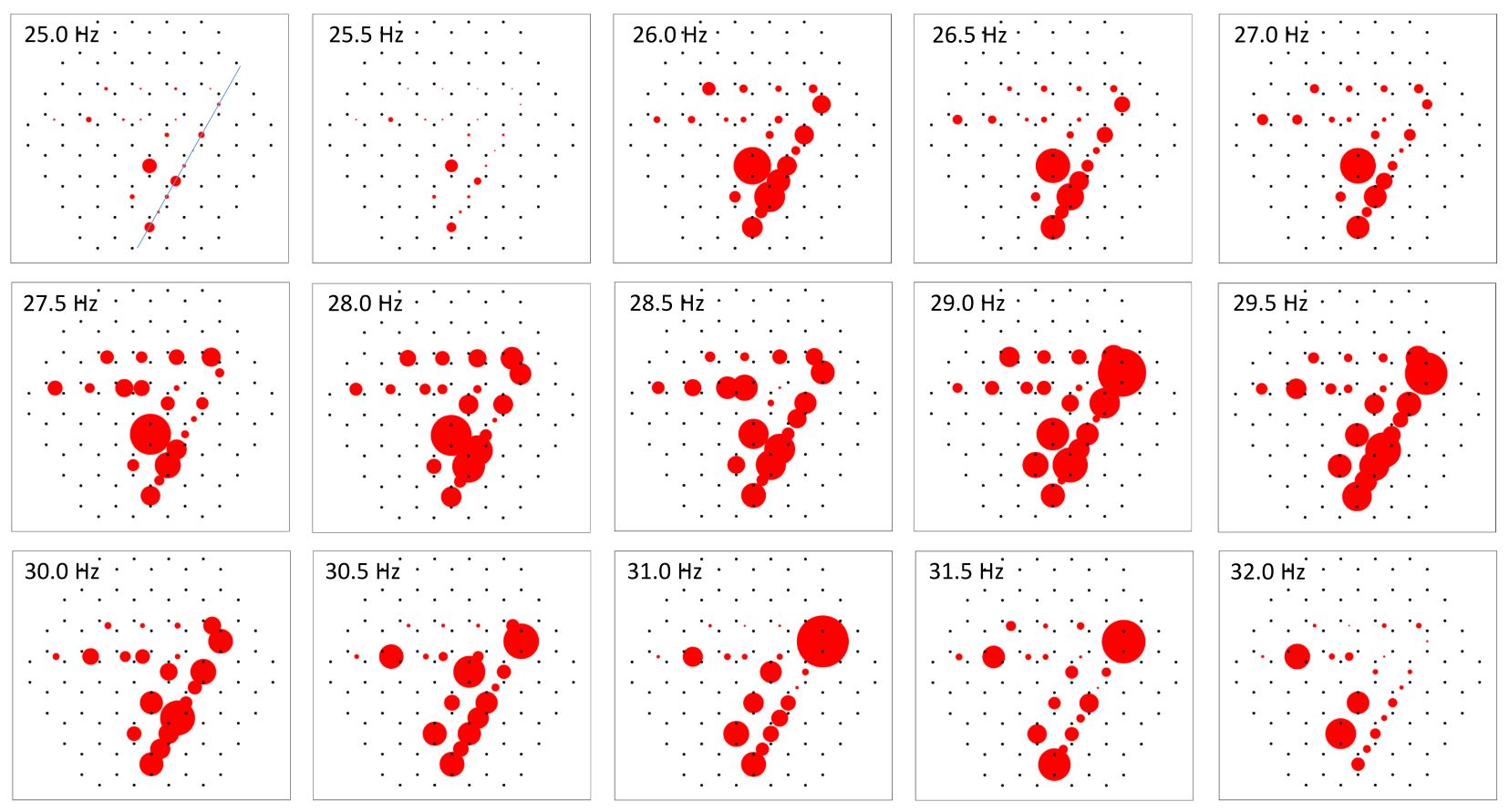}\\
  \caption{\small {Experimental observation of the L-shaped domain-wall modes.} Except for the shape of the interface, the rest of the details are as in Fig.~\ref{Fig:DataSEdge}.}
 \label{Fig:DataLEdge}
\end{figure*}

\vspace{0.2cm}

The results are reported in Fig.~\ref{Fig:DataSEdge}. One can see there that, for frequencies up to 25~Hz, the sensors return only small motion amplitudes. These frequencies must be within or very close to the bulk spectrum in which case the signal from the actuator disperses throughout the entire lattice, hence explaining the small amplitudes. Beyond 25~Hz, the sensors pick-up strong amplitudes near the interface and the amplitudes are seen to fade away into the bulk. We are definitely witnessing a standing wave supported by the interface channels. The strongest resonant patterns are observed within 28-29~Hz range of frequencies. Above this range, the sensor readings are seen again to fade away signaling that the frequency approaches the upper part of the bulk spectrum. Let us point out that these observations are in remarkable agreement with the data in Figs.~\ref{Fig:BulkThVsExp} and \ref{Fig:TheoreticalEdge}.

\vspace{0.2cm}

Our analysis assures us that the DW modes are robust against spatially slow deformations of the interface. However, it has been reported with many occasions \cite{LuPRL2016,LuNatPhys2017,PalNJP2017,VilaPRB2017,ZhuArxiv2017,JiangArxiv2017,LiuPRA2018,WuSciRep2018,ChaunsaliPRB2018,ChernArxiv2018}  that a signal can propagate along the domain-wall channels with very little back-reflection even if the interface is bent sharply. To investigate this interesting and potentially important effect, we reconfigured our system in the L-shaped DW configuration shown in Fig.~\ref{Fig:ExpLShapeWall}. The measurements have been repeated and the results are reported in Fig.~\ref{Fig:DataLEdge}. As many before us, we find that, indeed, there is a healthy transmission of the signal beyond the corner of the L-shaped DW. However, judging by the amplitudes seen along the two arms of the L-shape (see especially the panels $f=28.5$~Hz and $f=29$~Hz), we concluded that the transmission is only about 90\%. Nevertheless, witnessing such guiding of a mechanical signal along such fractured interface is quite an experience, as the reader can convince himself by examining the video recordings associated with the data in Fig.~\ref{Fig:DataLEdge}, supplied in the next section. 

\vspace{0.2cm}

The versatility of our experimental platform enabled us to explore the QVHE phenomenon quite thoroughly. For example, by replacing the spinner of type B with the one of type C shown in Fig.~\ref{Fig:Spinner}, we achieved the value $1.4$ for the critical parameter $r$. At a first sight, the data from Fig.~\ref{Fig:TheoreticalEdge} will suggest one that QVHE will be more robust for this $r$ but, to our surprise, it was not. Particularly, it did not increase the transmission across the corner of the L-shaped DW. We have also experimented with various perturbations and boundary conditions, some of which are supplied in the next section. At the end, we learned quite a number of important lessons and we will list them one by one in our last chapter. Let us conclude this section by pointing that the magnetically coupled spinners, due to their simplicity and versatility, can be a very effective educational platform where both experts and new comers can learn about the topological boundary modes.
  

\subsection{Experimental Details and Media Files}
\label{Sec:ExpTechnique}

One issue we faced right from the beginning is the large size of a fully self-assembled system. An effective solution we found was to divide the base that supports the assemblage in smaller panels which interlock one in another like a jigsaw puzzle. Henceforth, we laser cut 6 mm thick acrylic panels and fitted them with holes such that, when assembled, the holes generate a honeycomb lattice. The ball bearings shown in Fig.~\ref{Fig:Spinner}(a) were secured in place using zinc carriage bolts and the distance from spinners to base was maintained uniformly throughout the system with washers and nuts. Such panels can be readily assembled and then taken apart for storage or transportation. For example, this practical feature enabled us to demonstrate the topological boundary modes in front of several audiences.

\vspace{0.2cm}

The particular configurations shown in Figs.~\ref{Fig:ExpStraightWall} and \ref{Fig:ExpLShapeWall} were realized with eight acrylic panels and ninety six spinners. The ball bearings were fitted with arms and coupled by neodymium disc magnets, which were secured at the ends of specific arms with super glue.  The distance between adjacent magnets (always of opposite polarity) was 7.0 mm. Different configurations of the spinners can be achieved as illustrated in Fig.~\ref{Fig:Spinner}. Prior to running an experiment, all ball bearings were lubricated with silicone oil and the spinner arms were checked for tightness. We found that the connection arms need to be actually glued into place (we used Loctite Threadlocker Red 271 glue).


\vspace{0.2cm}

The lattice was actuated by a Pasco WA-9857 string vibrator, whose arm was fitted with a neodymium disc magnet and placed next to a connecting spinner arm, with the distance between the two held at 7.0 mm. The actuator was computer controlled by a custom LabVIEW software that drives a Rigol DG1022 function generator. That signal was amplified by a Crown XLS 2502 power amplifier with gain set to 5. Frequency sweeps between 8.0 and 40.0 Hz with two different frequency step sizes (0.5 Hz and 0.1 Hz) were performed. 

\vspace{0.2cm}

The data was collected by commercially available induction coil sensors, which were placed perpendicular and symmetrically on top of the magnetic bonds between spinners. Special place-holders were 3d-printed and hole locations were laser-cut into the base to ensure that the sensors are always placed in the same geometry relative to the arms. These sensors generate a time-oscillatory output, proportional to the rate of variation of the magnetic flux through the pick-up coil. In turn, these rates are proportional to the speed of the arms. For each frequency, the outputs were recorded and their root mean squares were extracted. The latter are proportional with the amplitudes of oscillations of the spinners and were used to generate the plots in Fig.~\ref{Fig:DataSEdge} and \ref{Fig:DataLEdge}.

\vspace{0.2cm}

\begin{figure*}[t]
\centering
\begin{subfigure}{0.23\textwidth}
\includemovie[poster=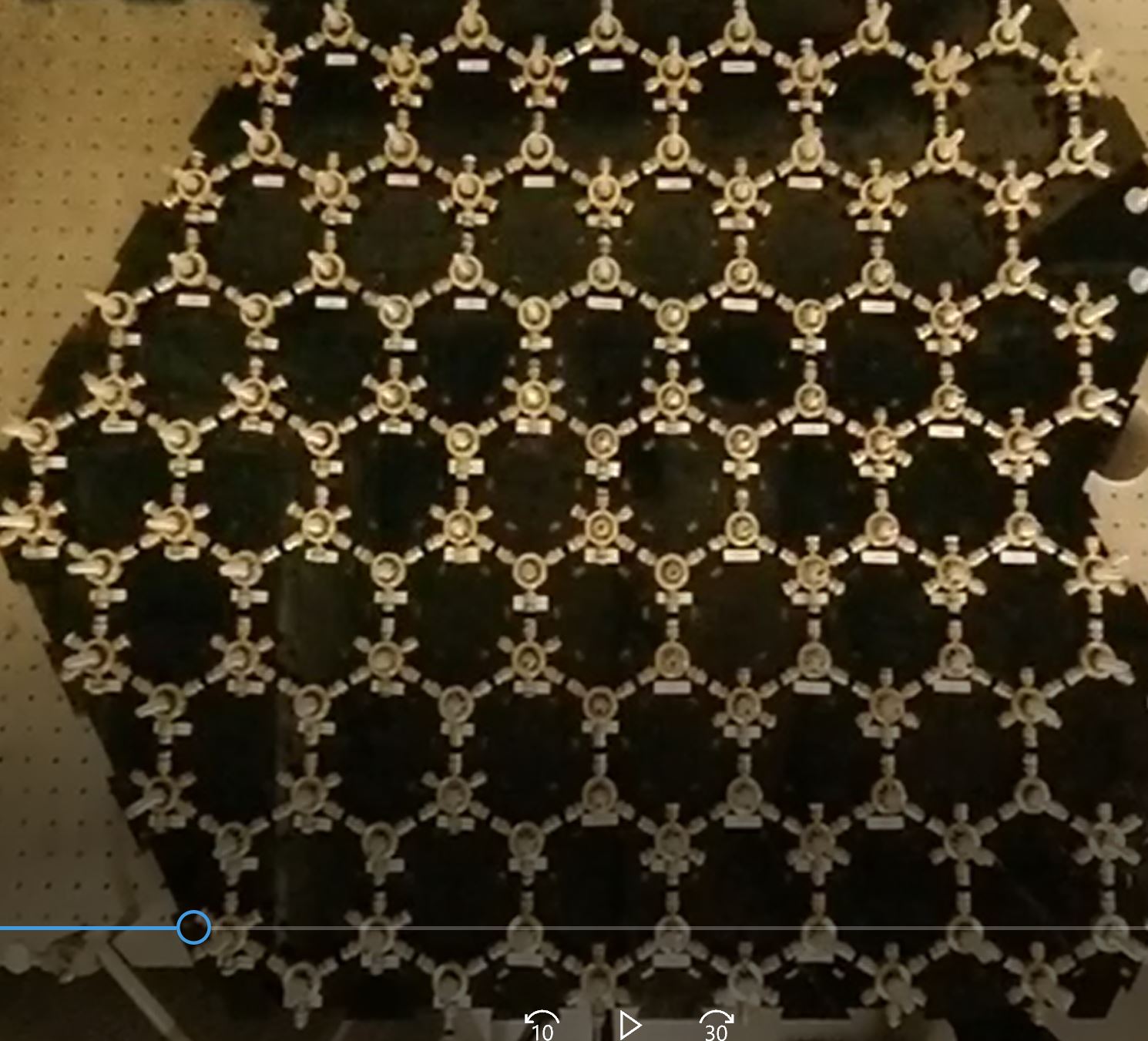,externalviewer,repeat=10,autoplay]{\linewidth}{\linewidth}{Bulk.gif}
\end{subfigure}
\begin{subfigure}{0.23\textwidth}
\includemovie[poster=Poster1.JPG,externalviewer,repeat=10,autoplay]{\linewidth}{\linewidth}{DWCleanTrim.gif}
\end{subfigure}
\begin{subfigure}{0.23\textwidth}
\includemovie[poster=Poster1.JPG,externalviewer,repeat=10,autoplay]{\linewidth}{\linewidth}{DWDefectLowerArmTrim.gif}
\end{subfigure}
\begin{subfigure}{0.23\textwidth}
\includemovie[poster=Poster1.JPG,externalviewer,repeat=10,autoplay]{\linewidth}{\linewidth}{DWDefectUpperArmTrim.gif}
\end{subfigure}
\caption{\small {\bf Video recordings of the experimental system.} (1st panel) Bulk resonant mode at frequency 16.3~Hz; (2nd panel) Clean L-shaped DW mode at frequency 28.5~Hz; (3rd panel) L-shaped DW mode with a defect in the lower arm at frequency 28.5~Hz; (4th panel) L-shaped DW mode with a defect in the upper arm at frequency 28.5~Hz. }
\label{Mov:AllMovies}
\end{figure*}

Video recording of the experimental system are reported in Fig.~\ref{Mov:AllMovies}, where the first panel show a bulk mode while the second panel show the DW mode corresponding to the panel at 28.5~Hz in Fig.~\ref{Fig:DataLEdge}. The remaining panels in Figs.~\ref{Mov:AllMovies} recorded the same mode but with one spinner removed in the lower/upper arm of the L-shaped DW.

\section{Conclusions} 
\label{Sec:Conclusions}

Our theoretical analysis provides a more complete framework to understand the topological nature of the QVHE, at both conceptual and quantitative levels. In this framework, for example, valley-Chern numbers can be defined in real-space and even in the presence of disorder. Furthermore, quantitative estimates about the faith of the effect become possible. In particular, a certain regime where QVHE occurs with full certainty was identified, giving guidance on how to engineer QVHE systems and keep them under control. This may be useful when one deals with lattices other than honeycomb, where QVHE has not been investigated so thoroughly. On the practical side, we became quite convinced that QVHE is robust in certain conditions that cover many situations of practical relevance. We believe these conditions can be controlled to a degree where QVHE becomes feasible for commercial applications.  


\vspace{0.2cm}

We learned a number of important lessons from our study. For example, while trying to understand why QVHE is more robust for $r=1.525$ than for $r=1.4$, we came to the conclusion that there are two factors here. First, there will be always a trade-off between the shape of the edge bands and the bulk gap. At $r=1.4$ the latter was smaller and, as a consequence, the domain-wall channels were spatially less localized. This is an important factor one needs to keep in mind when designing QVHE devices. The second factor was, to our surprise, that the honeycomb lattice is not optimal for the QVHE effect. Indeed, the valleys are located at $k_\| = \pm \frac{2\pi}{3}$ but, ideally, they will be located at $k_\|=\pm \frac{\pi}{2}$.  As a result, in the plots of the Berry curvature in Fig.~\ref{Fig:BerryCurvature}, one can see $F(\bm k)$ taking asymmetric values at $k_\|=0$ and $k_\|=\pm \pi$, which is an unwanted effect. It is also at $k\pm \frac{\pi}{2}$ where the counter-propagating domain-wall modes will be most separated from each other and hence more protected against backscattering. Examining the domain-wall dispersions in Fig~\ref{Fig:TheoreticalEdge}, one can then understand why the domain-wall modes were found to be more robust at energies closer to the upper bulk band rather than in the middle of the bulk gap. This also explains why $r=1.525$ was more optimal than $r=1.4$, because for the latter the interface bands were very close to the bulk spectrum near $k_\|=\pm \frac{\pi}{2}$. Let us mentioned that in our experimental platform, we can adjust the position of the valleys towards the optimal ones by decreasing the strength of the magnetic bonds within the primitive cells, which will lead to a new $\gamma(\bm k) = g + e^{\imath k_1} + e^{\imath k_2}$ with $g < 1$, while still preserving the inversion symmetry. Note that the $k_\| = \pm \frac{\pi}{2}$ cannot be achieved because that will require extreme limits, such as the complete decoupling of the the pairs of spinners inside the primitive cells.

\vspace{0.2cm}

We have a prediction for bilayers. While most of the existing literature is focused on DW modes, it becomes quite apparent from our analysis that bilayers support topological edge modes, provided the boundary conditions are properly engineered. For our system of spinners, for example, we predict that a simple layering followed by a magnetic coupling between top-bottom spinners at the edge, of similar strength as the bulk couplings, will generate two counter-propagating quasi-chiral edge modes. 

\vspace{0.2cm}

While the domain-wall modes appear to defy sharp corners, they are fully back-reflected when a spinner at the interface is jammed or removed. This can be witnessed first hand in the video recordings provided in the previous section. This behavior is in stark contrast with the edge excitations reported in \cite{Nash201514495} where the edge excitations of the mechanical Chern insulator are seen to find new propagation paths when obstructions are imposed. The difference is that QVHE is only a weak topological effect. In practice we can cope with this un-avoidable fact in two ways. First, one should note that QVHE is robust against layering and that the interface channel can be multiplied by layering. This will prevent an accidental jamming of few spinners from fully blocking the energy flow. It will also improve the transmissions across sharp corners of the DW. The second way is to take full control of the system. For example, if the coupling magnets are replaced by electro-magnets, then the domain-wall can be easily reconfigured so that energy is efficiently transfered from actuator to any point of the lattice. If a spinner gets jammed somewhere, a new interface can be reconfigured by re-programming the electro-magnets. The practical implications of such possibility have been already weighted in the literature \cite{ChengNatMat2016}.

\vspace{0.2cm}

This brings us to our concluding remark that, perhaps, the most important attribute of the QVHE is not in the topological protection of the DW modes but in the topological principles which enable one to design and mold them by only acting on space symmetries.  

\acknowledgments{ All authors acknowledge support from the W.M. Keck Foundation.}

\end{document}